\documentclass[%
 aip,
 amsmath,amssymb,
]{revtex4-1}
\usepackage{amsthm}
\usepackage{graphicx}% Include figure files
\usepackage{bm}% bold math
\usepackage[utf8]{inputenc}
\usepackage[T1]{fontenc}
\usepackage{mathptmx}
\usepackage{etoolbox}
\usepackage[normalem]{ulem} 
\usepackage{cancel}

\usepackage[shortlabels]{enumitem}

\usepackage{amsfonts}
\usepackage[utf8]{inputenc}
\usepackage[T1]{fontenc}
\usepackage{mathptmx}

\usepackage{color}

\makeatletter
\def\@email#1#2{%
 \endgroup
 \patchcmd{\titleblock@produce}
  {\frontmatter@RRAPformat}
  {\frontmatter@RRAPformat{\produce@RRAP{*#1\href{mailto:#2}{#2}}}\frontmatter@RRAPformat}
  {}{}
}%
\makeatother
\begin{document}

\preprint{AIP/123-QED}

\title{Persistent effects of inertia on diffusion-influenced reactions: Theoretical methods and applications \\
} %Title of paper

\author{Sangyoub Lee}
\email{sangyoub@snu.ac.kr}
\affiliation{Department of Chemistry, Seoul National University, Seoul 08826, South Korea}
%0000-0003-0577-3195

\author{Sergey D. Traytak}
\email{sergtray@mail.ru}
\affiliation{Semenov Federal Research Center for Chemical Physics, Russian Academy of Sciences, 4 Kosygina St., 119991 Moscow, Russian Federation}
%0000-0002-3551-8796

\author{Kazuhiko Seki}
\email[Author to whom correspondence should be addressed:]{k-seki@aist.go.jp}
\affiliation{National Institute of Advanced Industrial Science and Technology (AIST), Onogawa 16-1 AIST West, Ibaraki, 305-8569, Japan
}

\date{\today}% It is always \today, today,

\begin{abstract}
The Cattaneo--Vernotte model has been widely studied to take momentum relaxation into account in transport equations. 
Yet, the effect of reactions on the Cattaneo--Vernotte model has not been fully elucidated. 
At present, it is unclear how the current density associated with reactions can be expressed in the Cattaneo--Vernotte model. 
Herein, we derive a modified Cattaneo--Vernotte model by applying the projection operator method to the Fokker--Planck--Kramers equation with a reaction sink. 
The same modified Cattaneo--Vernotte model can be derived by a  Grad procedure.
We show that the inertial effect influences the reaction rate coefficient differently 
depending on whether the intrinsic reaction rate constant in the reaction sink term depends on the solute relative velocity or not.
The momentum relaxation effect can be expressed by a modified Smoluchowski equation including a memory kernel using the Cattaneo--Vernotte model. 
When the intrinsic reaction rate constant is independent of the reactant velocity and is localized, the modified Smoluchowski equation should be generalized to include a reaction term without a memory kernel. 
When the intrinsic reaction rate constant depends on the relative velocity of reactants, an additional reaction term with a memory kernel is required because of competition between the current density associated with the reaction and the diffusive flux during momentum relaxation. 
The competition effect influences even the long-time reaction rate coefficient. 

\end{abstract}

\maketitle

%%%%%%%%%%%%%%%%%%%%%%%%%%%%%%%%%%
%\section{Motivation}
%%%%%%%%%%%%%%%%%%%%%%%%%%%%%%%%%%
%\label{sec:motivation}

%%%%%%%%%%%%%%%%%%%%%%%%%%%%%%%%%%
%\section{Introduction}
%%%%%%%%%%%%%%%%%%%%%%%%%%%%%%%%%%
%\label{sec:intro}
%
%{\clr It is common knowledge that from a mathematical point of
%view, the theory of diffusion transfer is the counterpart of the heat
%transfer theory.
%
%
%bbbbbb}

%In physics, chemistry and biology, there are many processes occurring in ambient inert media that involve diffusion of small "guest" particles towards the surface of much larger ones and react with them after the contact.

% 
% It is presently accepted that for the most part, chemical reactions in micro-heterogeneous liquid media are {\it contact diffusion-influenced}. The latter means that the rate of these reactions is determined significantly by the rate of encounter
% of reactants due to diffusion; that is, the reaction rate is not controlled by only the chemical requirement of overcoming an activation energy barrier. Due to their high abundance, this type of reactions play a decisive role for a wide and diversified range of applications often occurring in physics, chemistry, biology and nanotechnology. Examples include excitation quenching of donors by acceptors, heterogeneous catalysis, crystal growth, Ostwald ripening, particle coagulation, crystal defect annealing, nutrient consumption by cells to name just a few.  
\section{Introduction}
\label{sec:I}

In the conventional theory of diffusion-influenced reactions,\cite{Rice_85} diffusive motion under a potential is expressed by the Smoluchowski equation and the reaction is taken into account by setting proper boundary conditions or adding a reaction sink term to the Smoluchowski equation. \cite{Rice_85} The latter approach is inevitable for taking into account long-range reactions, such as energy-transfer, and electron transfer in the Marcus inverted region. \cite{Wilemski_73,Murata_96,Seki_99}
For localized reactions, the effect of the reaction can be taken into account either by setting a boundary condition or introducing a reaction sink term with the reflecting boundary condition at the contact distance. \cite{COLLINS_49,SHOUP_82}
Although the conventional approach has been successful for studying most reactions in condensed phases, the more detailed motion of solutes should be taken into account to study reactions at shorter timescales compared with the solute velocity correlation time or to study collision-induced reactions. \cite{Traytak_23,Lee_23}

The most straightforward method to assess the short time and/or the effect of collision-induced reactions can be to perform Langevin dynamics simulations, molecular dynamics simulations, or Monte-Carlo simulations. \cite{Tachiya_86,Dong_89,Zhou_91,VanBeijeren_01,Yang_01,Litniewski_04,Lee_04,Kim_09,Piazza_13} 
For analytical approaches, reactions under ballistic transport at short times and diffusive transport at long times can be most simply formulated using the Fokker--Planck--Kramers equation (i.e., the Fokker--Planck equation in phase space). \cite{KRAMERS_40,Wang_45,DOI_75,Northrup_78,Naqvi_82,Harris_83,Naqvi_83,MOLSKI_88,Ibuki_97,Ibuki_03,Ibuki_06,Kim_09}
The formulation using the Fokker--Planck--Kramers equation is inevitable when a reaction is induced by collisions, where the intrinsic reaction rate constant for the Fokker--Planck--Kramers equation depends on the relative velocity of the reactants. \cite{DOI_75,Northrup_78,Naqvi_82,Harris_83,Naqvi_83,MOLSKI_88,Ibuki_97,Ibuki_03,Ibuki_06,Kim_09}
However, setting the boundary conditions to describe reactions in the Fokker--Planck--Kramers equation is not straightforward. \cite{Harris_78,Harris_81,Harris_82,Harris_83,Naqvi_82,Naqvi_83,MOLSKI_88,Ibuki_97,Ibuki_03}
Here, we note that a collision-induced reaction can be taken into account by adding the delta-function reaction sink term into the Fokker--Planck--Kramers equation.\cite{Northrup_78}
Systematic perturbation expansion in terms of the reaction sink term can be carried out. 

We present two approaches to eliminating velocity variables from the Fokker--Planck--Kramers equation with a reaction sink term. 
One approach is based on decoupling between the configurational distribution and momentum distribution, where an equilibrium Maxwell (Gaussian) distribution is assumed for the momentum. 
The second approach is based on the projection operator method with perturbation expansion. 
We show that the both approaches lead to the same modified Smoluchowski equation that includes a memory kernel and reaction sink terms. 
Using the decoupling approach, we also obtain differential equations for the density and current density. 
The differential equation for the current density can be regarded as the modified Cattaneo--Vernotte equation, where the inertial effect is modeled as a Markovian relaxation of the current density. 
The Cattaneo--Vernotte equation was modified, with an additional term added to account for collision-induced reactions. 

In the absence of a potential, the standard Cattaneo--Vernotte equation \cite{Cattaneo_58,VERNOTTE_58,Kubobook} reads  
\begin{align}
	\frac{\partial}{\partial t} \bm{j}(\bm{r} ,t)&=-\frac{1}{\tau_{\rm D}} 
	\left[ \bm{j}(\bm{r} ,t)+
	D \frac{\partial}{\partial \bm{r}} \rho(\bm{r},t)
	\right] . 
	\label{eq:Cattaneo_Intro1}
\end{align} 
Hereafter, $\rho(\bm{r},t)$ denotes the non-equilibrium pair correlation function (the concentration field of one reactant species at the position $\bm{r}$ at time $t$ around a reactant partner at the coordinate origin divided by the bulk concentration value) and $\bm{j}(\bm{r} ,t)$
is the associated current density. 
Corresponding arguments are the mutual separation vector $\bm{r}$ and time $t$ given in the augmented configuration space $ {\Bbb R}_{\bm{r}}^3 \times \lbrace t>0 \rbrace $; $\tau_{\rm D}$ and $D$ denote the momentum relaxation time and the mutual diffusion constant, respectively.

We show that 
the Cattaneo--Vernotte equation must be modified by the coupling with reaction as, 
\begin{align}
	\frac{\partial}{\partial t} \bm{j}(\bm{r} ,t)&=-\frac{1}{\tau_{\rm D}} 
	\left[ \bm{j}(\bm{r} ,t)+
	D \frac{\partial}{\partial \bm{r}}  \rho(\bm{r},t)-\tau_{\rm D} {\bf{R}}_{\rm v}(\bm{r},t)
	\right] , 
	\label{eq:Cattaneo_Intro2}
\end{align}
where 
the explicit expression for the term  
$\tau_{\rm D}{\bf{R}}_{\rm v}(\bm{r},t)$,  
representing the influence of a collision-induced reaction on the relaxation of the current density, will be given below.

The modification of the Cattaneo--Vernotte equation 
may be ignored when the intrinsic reaction rate constant in the reaction sink term of the Fokker--Planck--Kramers equation is independent of the solute relative velocity. 
Casting into the form of the Cattaneo--Vernotte equation clarifies physical interpretations of the effect of a collision-induced reaction on the current density. 
In addition to the Cattaneo--Vernotte equation, we show that the continuity equation is modified by a reaction sink term, 
\begin{align}
	\frac{\partial}{\partial t}  \rho(\bm{r},t)&=-\frac{\partial}{\partial \bm{r}} \cdot \bm{j}(\bm{r},t)
	-R (\bm{r},t), 
	\label{eq:Cattaneo_Intro3}
\end{align}
where the last term indicates 
an influence due to the reaction rate $R (\bm{r},t)$. 

Note that in our previous paper, \cite{Lee_23} we called the system of Eqs. (\ref{eq:Cattaneo_Intro2}) and (\ref{eq:Cattaneo_Intro3}) "diffusive Cattaneo system". 
It is 
noteworthy that Eq. (\ref{eq:Cattaneo_Intro3}) is an exact equation representing the conservation law, whereas
the standard Cattaneo--Vernotte equation [Eq. (\ref{eq:Cattaneo_Intro1})]
is essentially a constitutive equation, being an approximation based on 
 linear relaxation to local stationarity. 
Performing the known Kac's trick with the Cattaneo--Vernotte system of Eqs. (\ref{eq:Cattaneo_Intro2}) and (\ref{eq:Cattaneo_Intro3}), \cite{Kac_74} 
differentiating Eq. (\ref{eq:Cattaneo_Intro3}) with respect to $t$ and
applying the divergence operator to Eq. (\ref{eq:Cattaneo_Intro2}),
	we can eliminate $\bm{j}(\bm{r},t) $ and derive 
	the so-called telegraph equation for $ \rho(\bm{r},t) $ alone:
\begin{align}
	\tau_{\rm D}\frac{\partial^2}{\partial t^2} \rho(\bm{r},t)+ \frac{\partial}{\partial t}  \rho(\bm{r},t)&=
	D \frac{\partial^2}{\partial \bm{r}^2} \rho(\bm{r},t)
	-R (\bm{r},t)-\frac{\partial}{\partial t}R (\bm{r},t)-\tau_{\rm D} \frac{\partial}{\partial \bm{r}} \cdot {R}_{\rm v} (\bm{r},t). 
	\label{eq:Cattaneo_Intro4}
\end{align}
If the last two terms are not included, only one reaction sink term remains; the resultant equation has been called the hyperbolic reaction-diffusion equation. \cite{mendez_10,GHORAI_22,Traytak_23}
If the last term is dropped, the equation is called the reaction--telegraph equation and has been studied. \cite{mendez_10,Wakou_97,Holmes_93,Tilles_19,Lee_23}
We show that the reaction--telegraph equation is appropriate when the intrinsic reaction rate constant in the reaction sink term of the Fokker--Planck--Kramers equation is independent of reactant velocity, such as in cases of electron transfer and energy transfer. 
Our derivation can be regarded as a microscopic approach in which a reaction at the 
contact distance is considered. 
The reaction--telegraph equation has been studied in both mesoscopic and microscopic approaches. \cite{mendez_10,Wakou_97,Holmes_93,Tilles_19,Lee_23}
We emphasize that the last term in Eq. (\ref{eq:Cattaneo_Intro4}) is inevitable for collision-induced reactions and influences the long-time rate coefficient.

%%%%%%%%%%%%%%%%%%%%%%%%%%%%%%%%%%%%%%%%%%%%%%%%%%%%%%%%
%%%%%%%%%%%%%%%%%%%%%%%%%%%%%%%%%%%%%%%%%%%%%%%%%%%%%%%%
%\label{}
\section{Fokker--Planck--Kramers equation with a reaction sink term}
\label{sec:II}

Assume that the origin of the 
coordinates is located at the center of a reactant and $\sigma$ indicates a contact reaction distance. 
Let us consider the generalized non-equilibrium pair correlation function $f(\bm{r}, \bm{v},t)$, where $\bm{r}$ indicates the relative configuration vector between reactants and $\bm{v}$ 
stands for the relative velocity. Thus the function $f(\bm{r}, \bm{v},t)$ is defined in the augmented phase space $ (\bm{r}, \bm{v}, t) \in  \lbrace \bm{r} \in {\Bbb R}_{\bm{r}}^3: |\bm{r}|> \sigma \rbrace \times {\Bbb R}_{\bm{v}}^3 \times \lbrace t>0 \rbrace $.
The Fokker--Planck--Kramers equation with a reaction sink term introduced to 
incorporate the effects of a reaction
reads as 
\begin{align}
	 \frac{D}{Dt}f(\bm{r}, \bm{v},t) :=\frac{\partial}{\partial t}f(\bm{r}, \bm{v},t) + & \left(\bm{v} \cdot \frac{\partial}{\partial \bm{r}} - \frac{1}{\mu}  \frac{\partial U}{\partial \bm{r}} \cdot\frac{\partial}{\partial \bm{v}} \right)f(\bm{r}, \bm{v},t)  &  
	\nonumber \\
	& = \left[\frac{\partial}{\partial \bm{v}} \frac{1}{\tau_{\rm D}} \cdot \left(\bm{v} +\frac{k_{\rm B} T}{\mu} \frac{\partial}{\partial \bm{v}}
	\right)
	\right] f(\bm{r}, \bm{v},t) - R(\bm{r},\bm{v},t)\,, 
	\label{eq:1}
\end{align}
where the Fokker--Planck--Kramers operator is separated into a streaming and a collisional part in parentheses on 
the left hand side and in square brackets on the right hand side, respectively. $D/D t  $ denotes the material derivative.
For the subsequent study it is convenient to rewrite Eq. (\ref{eq:1}) as follows:
\begin{align}
	\frac{\partial}{\partial t} f(\bm{r}, \bm{v},t)=({\cal L}_0 +{\cal L}_1)f(\bm{r}, \bm{v},t), 
	\label{eq:1pro}
\end{align}
where ${\cal L}_0$ is the Fokker-Planck collision operator, 
\begin{align}
	{\cal L}_0 &= \frac{\partial}{\partial \bm{v}} \frac{1}{\tau_{\rm D}} \cdot \left(\bm{v} +\frac{k_{\rm B} T}{\mu} \frac{\partial}{\partial \bm{v}} \right)
	\label{eq:L0}
\end{align}	
and ${\cal L}_1$ is defined by 
\begin{align}
	{\cal L}_1 &= {\cal L}_{\rm L}- R(\bm{r},\bm{v},t), 
	\label{eq:L1}
\end{align}
using ${\cal L}_{\rm L}$ representing the streaming term:
\begin{align}
	{\cal L}_{\rm L} &= -\bm{v} \cdot\frac{\partial}{\partial \bm{r}}  +\frac{1}{\mu} \frac{\partial U}{\partial \bm{r}} \cdot \frac{\partial}{\partial \bm{v}} .
	\label{eq:Lv}
	\end{align}
In Eq. (\ref{eq:1}), $U (\bm{r})$ is an interacting potential, $\tau_{\rm D} =\mu/\xi_r$ is the momentum relaxation time constant, where $\mu$ and $ \xi_r $ are the reduced mass and the reduced friction coefficient, respectively \cite{Lee_23}, $k_{\rm B}$ is the Boltzmann constant, $T$ is the absolute temperature; 
$R(\bm{v},\bm{r},t)$ is the reaction sink term, 
whose explicit expression is given below. 
We consider 
isotropic systems, and introduce the unit normal vector to the reaction surface, $\bm{n} =\bm{r}/r$. 

Consider the case when the reaction does not occur; i.e. in Eq. (\ref{eq:1}) $R(\bm{r},\bm{v},t) \equiv 0$.
Posing 
the reflecting boundary condition for Eq. (\ref{eq:1}) we integrate it over $\bm{v}$. In this way the continuity equation is obtained by assuming that $f(\bm{r}, \bm{v},t)$ tends to zero exponentially
fast  
as $|\bm{v}|\rightarrow \infty$. 
As shown in Appendix A, using the divergence  
theorem for the continuity equation, Eq. (\ref{eq:1}) should be supplemented with a reflecting boundary condition expressed by
\begin{align}
	\int d\bm{v} \bm{n} \cdot \bm{v}f(\sigma-\epsilon, \bm{v},t) &=0,  
	\label{eq:2}
\end{align}
when the velocity components are integrated out, where 
$\sigma$ indicates a contact distance and 
$\epsilon$ indicates a small positive value; later, we take the limit of $\epsilon \rightarrow 0$. 
$\epsilon>0$ is introduced to avoid interference between reflection and reaction sink term represented using a delta-function set at $\sigma$ as shown below. 
The reflecting boundary is set at the radius $\sigma-\epsilon$ to ensure that not a half but the whole magnitude of the reactive sink term 
can be taken into account by the delta-function set at $\sigma$; 
the reflecting boundary condition approaches to $\sigma$ by taking the limit of $\epsilon \rightarrow 0$. 
Hence, 
$f(\bm{r}, \bm{v},t)$ is conserved by setting the reflecting boundary condition when $R(\bm{r},\bm{v},t) \equiv 0$ and, conversely, the reaction term breaks the conservation of 
$f(\bm{r}, \bm{v},t)$. 

For collision-induced reactions without reflection at a contact distance $\sigma$ in isotropic systems, 
the reaction sink term can be written as 
\begin{align}
	R(\bm{r}, \bm{v},t)=\frac{R_{\rm i}(r,\bm{v},t)}{4\pi r^2}\delta(r-\sigma) .
	\label{eq:bc1}
\end{align}
$R_{\rm i}(r,\bm{v},t)$ is given by \cite{MOLSKI_88}
\begin{align}
	R_{\rm i}(r,\bm{v},t)=-4\pi \sigma^2\bm{n} \cdot 
	\bm{v}\theta(-\bm{n} \cdot \bm{v}) f(r, \bm{v},t) ,
	\label{eq:bc1_1}
\end{align}
where $\theta(x)$ is the Heaviside step function;   
$\theta(x)=1$ for $x\geq 0$ otherwise zero. 
Equation (\ref{eq:bc1_1}) indicates that inward fluxes characterized by $v_z=\bm{n} \cdot \bm{v}\leq 0$ are perfectly absorbed. 
Here, the perfectly absorbing boundary condition is defined in phase space, 
which should be distinguished from the perfectly absorbing boundary condition in configuration space; 
the density at $\sigma$ is zero by imposing the perfectly absorbing boundary condition in configuration space. \cite{Rice_85}
In phase space, we need to take into account the ultimate escape from collision if the condition given by $v_z> 0$ is satisfied. 

For collision-induced reactions with a partially absorbing boundary condition at a contact distance $\sigma$, 
we introduce the fraction of reactive flux $f_{\rm r}(|v_z|)$ with $0<f_{\rm r} (|v_z|)\leq 1$. 
Equation (\ref{eq:bc1_1}) should be generalized to 
\begin{align}
	R_{\rm i}(r,\bm{v},t)=-4\pi \sigma^2\bm{n} \cdot 
	\bm{v}f_{\rm r} (|v_z|)\theta(-\bm{n} \cdot \bm{v}) f(r, \bm{v},t) .
	\label{eq:bc2}
\end{align}
The boundary condition for collision-induced reaction without reflection can be obtained when $f_{\rm r}(|v_z|)=1$. 
When $R_{\rm i}(r,\bm{v},t)$ is independent of $\bm{v}$, Eq. (\ref{eq:1}) posed earlier is applied; this equation has been studied explicitly for long-range reactions. \cite{DOI_75} 

%%%%%%%%%%%%%%%%%%%%%%%%%%%%%%%%%%%%%%%%%%%%%%%%%%%%%%%%%%%%%%
\section{Cattaneo--Vernotte model}
\label{sec:III}

Here we shall introduce the Cattaneo--Vernotte model by defining 
$\rho(\bm{r},t)$ and $\bm{j}(\bm{r},t)$
by the zeroth and first moments of the infinite moment equations for $ f(\bm{r}, \bm{v},t) $ \cite{Davies_54,Wilemski76} 
\begin{align}
	\rho(\bm{r},t)&=\int d\bm{v} f(\bm{r}, \bm{v},t) ,
	\label{eq:6d}\\
	\bm{j}(\bm{r},t) &=\int d\bm{v} \bm{v}f(\bm{r}, \bm{v},t), 
	\label{eq:flux}
\end{align}
respectively. 
Although only the first two moments are considered, our method belongs to the Grad procedure, 
where the orthogonality relation of Hermite functions is used as a moment closure method to an infinite hierarchy of kinetic equations for the Hermite moments;~\cite{Meyer83,Grad_49}
our results can be systematically extended to higher orders. 
Besides, the other systematic method may also be applicable here.~\cite{Traytak_14}
Using Eq. (\ref{eq:1}), we find
\begin{align}
	\frac{\partial}{\partial t} \rho(\bm{r},t)+\frac{\partial}{\partial \bm{r}} \cdot \bm{j}(\bm{r},t)=- \int d\bm{v} R(\bm{r},\bm{v},t)
	.
	\label{eq:2_7}
\end{align}
By multiplying $\bm{v}$ to Eq. (\ref{eq:1}) and integrating over $\bm{v}$, we obtain 
\begin{align}
	\frac{\partial}{\partial t} \bm{j}(\bm{r},t)&=-\frac{1}{\tau_{\rm D}} 
	\left[ \bm{j}(\bm{r},t)+\tau_{\rm D}
	\left(\int d\bm{v} \bm{v} \frac{\partial}{\partial \bm{r}} \cdot \bm{v}f(\bm{r}, \bm{v},t)
	+\frac{1}{\mu} \frac{\partial U}{\partial \bm{r}}\rho(\bm{r},t)+\int d\bm{v} \bm{v}R(\bm{r},\bm{v},t)\right) \right] .
	\label{eq:2_8}
\end{align}
For isotropic systems, the reflecting boundary condition can be written in terms of $\bm{j}(\bm{r},t)$ as 
\begin{align}
	\bm{n} \cdot \bm{j}(\sigma-\epsilon,t)&=0,   
	\label{eq:2_8BC}
\end{align}
from Eq. (\ref{eq:2}). 

We define the diffusion coefficient [$D_{ij}(t)$] using, \cite{Davies_54}
\begin{align}
	\rho(\bm{r},t) D_{ij}(t)=\tau_{\rm D}\int d\bm{v} 
	\bm{v}_i \bm{v}_j f(\bm{r}, \bm{v},t) , 
	\label{eq:2_10}
\end{align}
where $\bm{v}_i=\bm{v} \cdot \bm{r}_i/|\bm{r}_i|$ and $\bm{v}_j$ is defined in the similar manner. 
Equation (\ref{eq:2_8}) can be rewritten as 
\begin{align}
	\frac{\partial}{\partial t} \bm{j}_i(\bm{r},t)&=-\frac{1}{\tau_{\rm D}} 
	\left[ \bm{j}_i (\bm{r},t)+
	\sum_j \frac{\partial}{\partial \bm{r}_j} D_{ij}(t) \rho(\bm{r},t)
	+\frac{\tau_{\rm D}}{\mu} \frac{\partial U}{\partial \bm{r}_i}\rho(\bm{r},t)+\tau_{\rm D}\int d\bm{v} \bm{v}_i R(\bm{r},\bm{v},t)\right] .
	\label{eq:2_11}
\end{align}

When the velocity distribution is in equilibrium at the initial time, we introduce a decoupling approximation, 
\begin{align}
	f(\bm{r}, \bm{v},t)=\rho(\bm{r},t)g_{\rm eq}(\bm{v}) , 
	\label{eq:2_12}
\end{align}
where $g_{\rm eq}(\bm{v})$ indicates the Maxwell (Gaussian) distribution of velocity. 
Ignoring the hydrodynamic interactions effects, we can express the translational diffusion tensor [Eq. (\ref{eq:2_10})] using the Einstein relation as 
\begin{align}
	D_{ij}=\tau_{\rm D}\int d\bm{v} v_i v_j g_{\rm eq}(\bm{v}) = D \delta_{ij} , 
	\label{eq:2_13}
\end{align}
where $$D = \tau_{\rm D} \frac{k_{\rm B} T}{\mu}\,. $$ 
Thus, Eq. (\ref{eq:2_11}) can be simplified as
\begin{align}
	\frac{\partial}{\partial t} \bm{j}(\bm{r},t)&=-\frac{1}{\tau_{\rm D}} 
	\left[ \bm{j} (\bm{r},t)+
	D\left(\frac{\partial}{\partial \bm{r}} \rho(\bm{r},t)
	+\frac{1}{k_{\rm B} T} \frac{\partial U}{\partial \bm{r}}\rho(\bm{r},t)\right)+\tau_{\rm D}\int d\bm{v} \bm{v} R(\bm{r},\bm{v},t) \right] .
	\label{eq:2_14}
\end{align}
Equations (\ref{eq:2_7}) and (\ref{eq:2_14}) constitute the Cattaneo--Vernotte differential model extended to include a reaction. 

For collision-induced reactions in isotropic systems, we obtain  
\begin{align}
	\frac{\partial}{\partial t} \rho(r,t)&+
	\frac{1}{r^2}\frac{\partial}{\partial r} r^2 j_{r}(r,t)=- \kappa \rho(r,t)
	\frac{\delta(r-\sigma)}{4\pi r^2}, 
	\label{eq:2_22}
	\\
	\frac{\partial}{\partial t} j_{\rm r}(r,t)&=-\frac{1}{\tau_{\rm D}} 
	\left[ j_{\rm r} (r,t)+
	D\left(\frac{\partial}{\partial r} \rho(r,t)
	+\frac{1}{k_{\rm B} T} \frac{\partial U}{\partial r}\rho(r,t)\right)-\tau_{\rm D}\kappa_{\rm r} \rho(r,t)
	\frac{\delta(r-\sigma)}{4\pi r^2}\right] , 
	\label{eq:2_23}
\end{align}
where $\kappa$ is defined by 
\begin{align}
	\kappa=-4\pi \sigma^2\int d\bm{v}\bm{n} \cdot 
	\bm{v}f_{\rm r} (|v_z|)\theta(-\bm{n} \cdot \bm{v}) g_{\rm eq}(\bm{v}) ,
	\label{eq:23}
\end{align}
and $\kappa_{\rm r}$ satisfies $
\vec{\bm{\kappa}}=\kappa_{\rm r} \bm{n}$, 
where $\vec{\bm{\kappa}}$ is defined by
\begin{align}
	\vec{\bm{\kappa}}=4\pi \sigma^2\int d\bm{v}\bm{v} \bm{n} \cdot 
	\bm{v}f_{\rm r} (|v_z|)\theta(-\bm{n} \cdot \bm{v}) g_{\rm eq}(\bm{v}) .
	\label{eq:24}
\end{align}
In Eq. (\ref{eq:2_14}), the term $\tau_{\rm D}\int d\bm{v} \bm{v} R(\bm{r},\bm{v},t)$ indicates the influence of a collision-induced reaction on the relaxation of the current density. 
The current density relaxes toward the steady-state quantity denoted by $\bm{j}_{\rm s} (\bm{r})$:
\begin{align}
	\bm{j}_{\rm s} (\bm{r})=-
	D\left(\frac{\partial}{\partial \bm{r}} \rho_{\rm s} (\bm{r})
	+\frac{1}{k_{\rm B} T} \frac{\partial U}{\partial \bm{r}}\rho_{\rm s} (\bm{r})\right)-\tau_{\rm D}\int d\bm{v} \bm{v} R_{\rm s} (\bm{r},\bm{v})  ,
	\label{eq:2_14_1}
\end{align}
where the subscript denoted by ``s'' indicates the steady-state quantity. 
The last term in Eq. (\ref{eq:2_14_1}) can be expressed as 
\begin{align}
	\tau_{\rm D}\int d\bm{v} \bm{v} R_{\rm s} (\bm{r},\bm{v}) =-\tau_{\rm D}\vec{\bm{\kappa}}\rho_{\rm s}(r)
	\frac{\delta(r-\sigma)}{4\pi r^2} , 
	\label{eq:Rst}
\end{align}
with $\vec{\bm{\kappa}}$ given by Eq. (\ref{eq:24}). 
$\vec{\bm{\kappa}}$ should be an outward vector at $\sigma$ when $\bm{v}$ is the inward velocity induced by a collision-induced reaction because $-\bm{n} \cdot \bm{v}$ should be negative to have a non-zero value because of $\theta(-\bm{n} \cdot \bm{v})$.   
Later, in Eq. (\ref{eq:30}), we show that $\kappa_{\rm r} >0$. 
The result $\kappa_{\rm r} >0$ indicates a positive correlation of the velocity vector and the inward normal component of the velocity vector at the contact distance.  
Equation (\ref{eq:2_14_1}) indicates that the inward current density is reduced by the positive correlation at the contact distance. 
In a steady state, the reduction of the inward current density is given by the last term in Eq. (\ref{eq:2_14_1}), which shows the outward current density. 
The sign of the last term in Eq. (\ref{eq:2_23}) is chosen accordingly. 

When $f_{\rm r} (|v_z|)=f_{\rm r}$, where $f_{\rm r}$ is a constant satisfying $0<f_{\rm r} \leq1$, Eq. (\ref{eq:23}) is simplified to  
\begin{align}
	\kappa=-4\pi \sigma^2\int d\bm{v}\bm{n} \cdot 
	\bm{v}f_{\rm r} \theta(-\bm{n} \cdot \bm{v}) g_{\rm eq}(\bm{v}) .
	\label{eq:25}
\end{align}
We choose the $z$-axis in the Cartesian coordinate for expressing $\bm{v}$ [$\bm{v}=(v_x, v_y, v_z)$] in the direction of $\bm{n}$. 
In this coordinate, we have $\bm{n} \cdot \bm{v}=v_z$ and $\int_{-\infty}^\infty d v_z \theta(-v_z) \cdots =\int_{-\infty}^0 d v_z \cdots $. 
Using the explicit expression
\begin{align}
	g_{\rm eq}(\bm{v})=\left(
	\frac{\mu}{2\pi k_{\rm B}  T}
	\right)^{3/2} \exp\left(-\frac{\mu v^2}{2k_{\rm B} T}\right) 
	\label{eq:27}
\end{align}
with $v^2=v_x^2+v_y^2+v_z^2$, we find using Eq. (\ref{eq:2_13}) \cite{Kapral_78,Naqvi_83,MOLSKI_88,Zhou_91,Kim_09} that
\begin{align}
	\kappa&=2\sigma^2\sqrt{\frac{2\pi k_{\rm B} T}{\mu}} f_{\rm r}=2\sigma^2  \sqrt{\frac{2\pi D}{\tau_{\rm D}}}\, f_{\rm r}. 
	\label{eq:28}
\end{align}

We also calculate $\vec{\bm{\kappa}}$:
\begin{align}
	\vec{\bm{\kappa}}=4\pi \sigma^2 f_{\rm r} \int d\bm{v} \bm{v} \left( \bm{n} \cdot \bm{v} \right)
	\theta(-\bm{n} \cdot \bm{v}) g_{\rm eq}(\bm{v}) .
	\label{eq:26}
\end{align}
When evaluating $\vec{\bm{\kappa}}$, we choose the $z$-axis in the Cartesian coordinates for expressing $\bm{v}$ in the direction of $\bm{n}$ and find $\vec{\bm{\kappa}}=\kappa_{\rm r} \bm{n}$ with
\begin{align}
	\kappa_{\rm r}=\frac{2 \pi \sigma^2 k_{\rm B} T}{\mu} f_{\rm r}=  2 \pi \sigma^2 \frac{D}{\tau_{\rm D}} f_{\rm r}, 
	\label{eq:30}
\end{align}
where Eq. (\ref{eq:2_13}) is used. 
Equation (\ref{eq:26}) can be interpreted as the correlation between the velocity vector and the inward normal component of the velocity vector at the contact distance.  
Equation (\ref{eq:30}) with Eq. (\ref{eq:2_23}) indicates that the correlation at the contact distance is positive and reduces the inward current density. 

The perfectly absorbing boundary condition for a collision-induced reaction implies that all inward fluxes lead to a reaction and that the reflected outward fluxes are zero at the reactive boundary. \cite{Harris_81,Burschka_81}
In this case, a half-Gaussian distribution, where the velocity component in the outward direction at the reactive boundary is set to zero, has been introduced. 
In general, the velocity vector projected to $\bm{n}$ in the negative direction is absorbed and the Gaussian distribution can be distorted.
Determining the proper distribution requires another approach. 
For the moment, we introduce the Gaussian distribution in Eq. (\ref{eq:2_12}). 
In the subsequent section, we apply the projection operator method to derive the corresponding terms perturbatively. 

%%%%%%%%%%%%%%%%%%%%%%%%%%%%%%%%%%%%%%%%%%%%%%%%%%%%%%%%%%%%%%%%%%%%%%%%
We denote the Laplace transform of $\rho(r,t)$ and $j_{\rm r}(r,t)$ with respect to $t$ by $\hat{\rho}(r,s)$ and $\hat{j}_{\rm r}(r,s)$, respectively. 
By the Laplace transform of Eq. (\ref{eq:2_23}), %
we obtain, 
\begin{align}
	\hat{j}_{\rm r}(r,s)-\tau_{\rm D}(s) j_{\rm r}(r,0)&=-
	\left[
	D(s)\left(\frac{\partial}{\partial r} \hat{\rho}(r,s)
	+\frac{1}{k_{\rm B} T} \frac{\partial U}{\partial r}\hat{\rho}(r,s)\right)-\tau_{\rm D}(s)  \kappa_{\rm r} \hat{\rho}(r,s) 
	\frac{\delta(r-\sigma)}{4\pi r^2}\right] ,
	\label{eq:2_23_L}
\end{align} 
where $\tau_{\rm D}(s)=\tau_{\rm D}/(1+s\tau_{\rm D})$ and $D(s)=D/(1+s\tau_{\rm D})$ are introduced. 
By substituting Eq. (\ref{eq:2_23_L}) into the Laplace transform of Eq. (\ref{eq:2_22}), 
we find 
\begin{align}
	s\hat{\rho}(r,s)-
	\rho(r,0)+
	\tau_{\rm D}(s) \frac{1}{r^2}\frac{\partial}{\partial r} r^2 
	j_{\rm r}(r,0)
	&=\frac{1}{r^2}\frac{\partial}{\partial r} D(s) r^2 
	\frac{\partial}{\partial r}
	\hat{\rho}(r,s)
	\nonumber \\
	&
	-\kappa \hat{\rho}(r,s)
	\frac{\delta(r-\sigma)}{4\pi r^2}-
	\tau_{\rm D}(s)  \frac{1}{r^2}\frac{\partial}{\partial r} r^2 \kappa_{\rm r}\hat{\rho}(r,s) \frac{\delta(r-\sigma)}{4\pi r^2} , 
	\label{eq:2_31CV}
\end{align}
where the potential function is set equal to zero for simplicity. 
%%%%%%%%%%%%%%%%%%%%%%%%%%%%%%%%%%%%%%%%%%%%%%%%%%%%%%%%%%%%%%%%%%%%%%%%%%%%%%%%%%%

Equations (\ref{eq:2_22}) and (\ref{eq:2_23}) can be expressed as the modified telegraph equation given by a function of $\rho(r,t)$ alone: 
\begin{align}
	\tau_{\rm D} &\frac{\partial^2}{\partial t^2} \rho(r,t)+  \frac{\partial}{\partial t} \rho(r,t)
	=\frac{1}{r^2}\frac{\partial}{\partial r} D r^2  
	\left(\frac{\partial}{\partial r} \rho(r,t)
	+\frac{1}{k_{\rm B} T} \frac{\partial U}{\partial r}\rho(r,t)\right)
	\nonumber \\
	&
	- \left(\kappa+\tau_{\rm D}  \frac{\partial}{\partial t}\kappa \right) \rho(r,t)
	\frac{\delta(r-\sigma)}{4\pi r^2}-
	\tau_{\rm D}  \frac{1}{r^2}\frac{\partial}{\partial r} r^2 \kappa_{\rm r}\rho(r,t) \frac{\delta(r-\sigma)}{4\pi r^2} , 
	\label{eq:2_28}
\end{align}
where the boundary condition must be given by \cite{Naqvi_82,Naqvi_83}
\begin{align}
	\left.\left(\frac{\partial}{\partial r} \rho(r,t)
	+\frac{1}{k_{\rm B} T} \frac{\partial U}{\partial r}\rho(r,t)\right) 
	\right|_{r=\sigma-\epsilon}&=0 .
	\label{eq:bc_2_26}
\end{align}
In the Laplace domain, Eq. (\ref{eq:2_28}) can be expressed as 
\begin{align}
	s\hat{\rho}(r,s)-
	\rho(r,0)-
		& \tau_{\rm D}(s) 
		\left[\left. \frac{\partial}{\partial t} \rho(r,t)\right|_{t=0} +\kappa \rho(r,0)
	\frac{\delta(r-\sigma)}{4\pi r^2} \right]
	=\frac{1}{r^2}\frac{\partial}{\partial r} D(s) r^2 
	\frac{\partial}{\partial r}
	\hat{\rho}(r,s)
	\nonumber \\
	&
	-\kappa \hat{\rho}(r,s)
	\frac{\delta(r-\sigma)}{4\pi r^2}-
	\tau_{\rm D}(s)  \frac{1}{r^2}\frac{\partial}{\partial r} r^2 \kappa_{\rm r}\hat{\rho}(r,s) \frac{\delta(r-\sigma)}{4\pi r^2} , 
	\label{eq:2_31T}
\end{align}
where $U=0$ is assumed and we take into account that the right-hand side of Eq. (\ref{eq:2_28}) contains the term  
\begin{align}
	- \tau_{\rm D}  \frac{\partial}{\partial t}\kappa \rho(r,t)
	\frac{\delta(r-\sigma)}{4\pi r^2} .
\nonumber 
\end{align}
We consider the case of the homogeneous initial conditions  
\begin{align}
	\rho(r,0) =H(r-\sigma)\,, \quad 
	\left. \frac{\partial}{\partial t} \rho(r,t)\right|_{t=0}=0\,, \quad 
	j_{\rm r}(r,0)=0\,, 
	\label{eq:bc_2_26a}
\end{align} 
where   
$H(x)=1$ for $x>0$ otherwise zero;
[$H(x)=1-\theta(-x)$].
%%%%%%%%%%%%%%%%%%%%%%%%%%%%%%%%%%%%%%%%%%%%%%%%%%%%%%%%%%%%%%%%%%%%%%%%%%%%%%%%%%%%%
When the initial conditions are given by Eq. (\ref{eq:bc_2_26a}), 
both Eqs. (\ref{eq:2_31CV}) and (\ref{eq:2_31T}) reduce to, 
\begin{align}
	s\hat{\rho}(r,s)-
	\rho(r,0)
	&=\frac{1}{r^2}\frac{\partial}{\partial r} D(s) r^2 
	\frac{\partial}{\partial r}
	\hat{\rho}(r,s)
	\nonumber \\
	&
	-\kappa \hat{\rho}(r,s)
	\frac{\delta(r-\sigma)}{4\pi r^2}-
	\tau_{\rm D}(s)  \frac{1}{r^2}\frac{\partial}{\partial r} r^2 \kappa_{\rm r}\hat{\rho}(r,s) \frac{\delta(r-\sigma)}{4\pi r^2} , 
	\label{eq:2_31}
\end{align}
which will be also derived using the projection operator method in the subsequent section.
Therefore, Eq. (\ref{eq:2_31}) can be derived either from Eqs. (\ref{eq:2_22}) and (\ref{eq:2_23}) (the Cattaneo--Vernotte differential model) with the initial condition given by $ j_{\rm r}(r,0)=0$ or 
from Eq. (\ref{eq:2_28}) (the modified telegraph equation) for the initial condition given by Eq. (\ref{eq:bc_2_26a}). 
It should be reminded that the first initial condition of Eq. (\ref{eq:bc_2_26a}) is not required to derive Eq. (\ref{eq:2_31}) from Eqs. (\ref{eq:2_22}) and (\ref{eq:2_23}) (the Cattaneo--Vernotte differential model);  
the initial condition for $\rho(r,0)$ can be chosen arbitrary. 
Moreover, 
the first initial condition of Eq. (\ref{eq:bc_2_26a}) should be given by $H(r-\sigma)$ rather than $\theta(r-\sigma)$ because of the 4th term in Eq. (\ref{eq:2_31T}). 
In these senses, 
it might be preferable to use the Cattaneo--Vernotte differential model rather than the modified telegraph equation as already concluded previously. \cite{Lee_23}
%%%%%%%%%%%%%%%%%%%%%%%%%%%%%%%%%%%%%%%%%%%%%%%%%%%%%%%%%%%%%%%%%%%%%%%%%%%%%%%%%%%%%%%%%%%%%%%%%%%%

By multiplying $\int_{\sigma-\epsilon}^{\infty} 4\pi r^2 dr$ on both sides of Eq. (\ref{eq:2_31}) and introducing 
$\hat{p}(s)=\int_{\sigma-\epsilon}^{\infty} 4\pi r^2 dr\hat{\rho}(r,s)$ with $p(0)=1$,  we obtain
\begin{align}
	s\hat{p}(s)-1
	&=
	-\kappa \hat{\rho}(\sigma,s), 
	\label{eq:2_32}
\end{align}
where the partial integration was performed to evaluate the last term associated with $\kappa_{\rm r}$ in Eq. (\ref{eq:2_31}). 
By applying the inverse Laplace transform, we obtain from Eq. (\ref{eq:2_32}), 
\begin{align}
\frac{d}{dt} p(t)=-\kappa  \rho(\sigma,t), 
\label{eq:2_32_1}
\end{align}
and the rate coefficient can be given by (see Appendix B), 
\begin{align}
k(t)=\kappa \rho(\sigma,t),  
\label{eq:ratec1}
\end{align}
which can be expressed in the Laplace domain as 
$\hat{k}(s)=\kappa  \hat{\rho}(\sigma,s)$.   
In principle, the reaction rate coefficient should 
be obtained by taking the limit of $\epsilon \rightarrow 0$ after the inverse Laplace transform of $\kappa \hat{\rho}(\sigma,s)$, 
where the reflecting boundary condition is set at $\sigma-\epsilon$. 
Here, we assume that the limits are exchangeable. 

It is instructive to express Eq. (\ref{eq:2_31}) in the time domain as 
\begin{align}
	\frac{\partial}{\partial t} \rho(r,t)&=
	\int_0^t d t_1 \exp\left(- \frac{t-t_1}{\tau_{\rm D}}\right) 
	\left[\frac{1}{r^2}\frac{\partial}{\partial r} \frac{D}{\tau_{\rm D}} r^2 
	\frac{\partial}{\partial r}\rho(r,t_1) -\frac{1}{r^2}\frac{\partial}{\partial r} r^2 \kappa_{\rm r}\rho(r,t_1) \frac{\delta(r-\sigma)}{4\pi r^2} 
	\right]
	\nonumber \\
	&
	-\kappa \rho(r,t)
	\frac{\delta(r-\sigma)}{4\pi r^2} .
	\label{eq:2_31_tm}
\end{align}
The term multiplied by $\kappa_{\rm r}$ is coupled to the memory kernel, whereas the term multiplied by $\kappa$ is not coupled to the memory kernel.
The term multiplied by $\kappa_{\rm r}$ indicates a reduction of the current density by the positive correlation between the velocity vector and the inward normal component of the velocity vector at the contact distance, as explained below Eq. (\ref{eq:Rst}).
The memory kernel multiplied to $\kappa_{\rm r}$ represents the competition of the current density associated with collision-induced reaction and the diffusive flux during momentum relaxation (i.e., an inertial effect).

%%%%%%%%%%%%%%%%%%%%%%%%%%%%%%%%%%%%%%%%%%%%%%%%%%%%%%%%%%%%%%%%%%%%%%%%%%%%%%%%%%%%%%%%
\section{Projection operator method}
\label{sec:IV}

Here, we apply the projection operator method to Eq. (\ref{eq:1pro}) to obtain a closed equation for $\rho(\bm{r},t)$. 
In Eq. (\ref{eq:1pro}),  an explicit expression of $R(\bm{r},\bm{v},t)$ is substituted in ${\cal L}_1$; 
using Eq. (\ref{eq:bc2}), 
we first consider 
${\cal L}_1= {\cal L}_{\rm L}+{\cal L}_{\rm R}$, where ${\cal L}_{\rm R} $ is given by, 
\begin{align}
	{\cal L}_{\rm R} &= \bm{n} \cdot 
	\bm{v} f_{\rm r} (|v_z|)\theta(-\bm{n} \cdot \bm{v}) \delta(r-\sigma).
	\label{eq:LR}
\end{align}
We define the projection operator for any function ${\cal O} (\bm{r},\bm{v})$ as \cite{Kubobook,Northrup_78,Seki_99,Bandyopadhyay_00}
\begin{eqnarray}
	{\cal P} {\cal O} (\bm{r},\bm{v}) = g_{\rm eq}(\bm{v} )  \int \mbox{d} \bm{v} 
	{\cal O}(\bm{r},\bm{v}),  
\end{eqnarray}
and the complementary projection operator by 
\begin{eqnarray}
	{\cal Q} = 1 - {\cal P} .
\end{eqnarray}
We note that ${\cal L}_0 g_{\rm eq}(\bm{v} )  =0$ and 
\begin{eqnarray}
	{\cal P} {\cal L}_0 \cdots &=& 0,  
	\label{eq:PL0}
\end{eqnarray}
where we used the fact that $g_{\rm eq}(\bm{v} )$ 
exponentially goes to zero as  $|\bm{v}_i| \rightarrow \infty$.   
Following the conventional projection operator formalism using ${\cal L}={\cal L}_0+{\cal L}_1$, we first introduce \cite{Kubobook}
\begin{eqnarray}
	\frac{\partial}{\partial t} {\cal{P}} f(\bm{r}, \bm{v},t) &=&   
	{\cal P} 
	({\cal L} {\cal P} f 
	+ {\cal L} {\cal Q} f)  ,
	\label{projection} \\
	\frac{\partial}{\partial t} {\cal Q} f(\bm{r}, \bm{v},t) &=& 
	{\cal Q}
	({\cal L} {\cal Q} f 
	+ {\cal L} {\cal P} f) .  
	\label{eqpdot}
\end{eqnarray}
The solution of Eq. (\ref{eqpdot}) can be formally expressed as 
\begin{align}
	{\cal Q} f(\bm{r}, \bm{v},t)=\int_0^t dt_1\exp\left[(t-t_1) {\cal Q} {\cal L} \right] {\cal Q}
	{\cal L} {\cal P} f(\bm{r}, \bm{v},t_1) +\exp\left(t {\cal Q} {\cal L} \right) {\cal Q} f_0 ,
	\label{eq:formalsolP'}
\end{align}
where $f_0=f(\bm{r}, \bm{v},0)$. 
By substituting Eq. (\ref{eq:formalsolP'}) into Eq. (\ref{projection}), we can express the formally closed equation 
${\cal{P}} f(\bm{r}, \bm{v},t)$ as
\begin{align}
	\frac{\partial}{\partial t} {\cal{P}} f(\bm{r}, \bm{v},t) =& {\cal P} {\cal L} {\cal P} f (\bm{r}, \bm{v},t) +{\cal P} {\cal L}\int_0^t dt_1\exp\left[(t-t_1) {\cal Q} {\cal L} \right] {\cal Q}
	{\cal L} {\cal P} f (\bm{r}, \bm{v},t_1)
	\nonumber \\
	&
	+{\cal P} {\cal L}\exp\left(t {\cal Q} {\cal L} \right) {\cal Q} f_0 . 
	\label{eq:formalsolP}
\end{align}

We consider the case of an initial equilibrium velocity distribution $f_0 =\rho(\bm{r}, 0)g_{\rm eq}(\bm{v} )$ and find ${\cal Q} f_0 =0$ because
${\cal P} g_{\rm eq}(\bm{v} ) = g_{\rm eq}(\bm{v} )$. 
By noting that ${\cal L}_0 {\cal P}=0$ using ${\cal L}_0 g_{\rm eq}(\bm{v} )  =0$ and Eq. (\ref{eq:PL0}), we can simplify Eq. (\ref{eq:formalsolP}) to 
\begin{align}
	\frac{\partial}{\partial t} {\cal{P}} f(\bm{r}, \bm{v},t) &= {\cal P} {\cal L}_1 {\cal P} f (\bm{r}, \bm{v},t) 
	+{\cal P} {\cal L}_1\int_0^t dt_1\exp\left[(t-t_1) {\cal Q} {\cal L} \right] {\cal Q}
	{\cal L}_1 {\cal P} f (\bm{r}, \bm{v},t_1)  . 
	\label{eq:formalsolP1}
\end{align}
Because $g_{\rm eq}(\bm{v} )$ 
exponentially goes to zero as $|\bm{v}_i| \rightarrow \infty$, we have 
\begin{align}
	{\cal P}\frac{\partial}{\partial \bm{v}} \cdots {\cal P} f =0;
\end{align}
we also have ${\cal{P}}\bm{v} \cdots {\cal P} f=0$. 
Equation (\ref{eq:formalsolP1}) is further simplified to 
\begin{align}
	\frac{\partial}{\partial t} {\cal{P}} f(\bm{r}, \bm{v},t) &= {\cal P} {\cal L}_{\rm R} {\cal P} f  +{\cal P}
	\left(- \bm{v} \cdot \frac{\partial}{\partial \bm{r}} +{\cal L}_{\rm R}  \right)\int_0^t dt_1\exp\left[(t-t_1) {\cal Q} {\cal L}\right] {\cal Q}
	{\cal L}_1 {\cal P} f (\bm{r}, \bm{v},t_1) . 
	\label{eq:formalsolP2}
\end{align}
By applying perturbation expansion for the reaction sink term and the streaming term, we obtain the lowest-order approximation by changing $(t-t_1) {\cal Q} {\cal L}$ to $(t-t_1) {\cal L}_0$, where Eq. (\ref{eq:PL0}) is used; we also obtain 
\begin{align}
	\frac{\partial}{\partial t} \rho(\bm{r}, t) &\approx \int \mbox{d} \bm{v}  {\cal L}_{\rm R} g_{\rm eq} \rho  +\int \mbox{d} \bm{v} 
	\left(- \bm{v} \cdot \frac{\partial}{\partial \bm{r}} +{\cal L}_{\rm R}  \right)\int_0^t dt_1\exp\left[(t-t_1) {\cal L}_0 \right] {\cal Q}
	{\cal L}_1 g_{\rm eq}  \rho(\bm{r},t_1)  , 
	\label{eq:formalsolP3}
\end{align}
where ${\cal P} f=g_{\rm eq} \rho$. 
Using Eq. (\ref{eq:23}), we can express 
$\int \mbox{d} \bm{v}  {\cal L}_{\rm R} g_{\rm eq}=-\kappa/(4\pi \sigma^2) \delta(r-\sigma)$ and 
\begin{align}
	\frac{\partial}{\partial t} \rho(\bm{r}, t) =& -\kappa  \rho(\bm{r}, t) \frac{\delta(r-\sigma)}{4\pi \sigma^2} +
	\nonumber \\
	&
	\int \mbox{d} \bm{v} 
	\left(- \bm{v} \cdot \frac{\partial}{\partial \bm{r}}+{\cal L}_{\rm R}  \right)\int_0^t dt_1\exp\left[(t-t_1) {\cal L}_0 \right] {\cal Q}
	{\cal L}_1 g_{\rm eq} \rho (\bm{r}, t_1) . 
	\label{eq:formalsolP3_1}
\end{align}

We here further study the last term in Eq. (\ref{eq:formalsolP3_1}). 
By introducing 
\begin{align}
	\frac{\partial}{\partial {\bm v}} g_{\rm eq}(\bm{v} )=-\frac{\mu {\bm v}}{k_{\rm B} T} g_{\rm eq}(\bm{v}) ,
	\label{eq:gdv}
\end{align}
we obtain
\begin{align}
	{\cal L}_{\rm L} g_{\rm eq}\rho(\bm{r}, t_1)=-{\bm v}\cdot \left(\frac{\partial}{\partial \bm{r}} + 
	\frac{1}{k_{\rm B} T} \frac{\partial U}{\partial \bm{r}} \right) g_{\rm eq}\rho (\bm{r}, t_1). 
	\label{eq:LLgeq}
\end{align}
Because ${\bm v}g_{\rm eq}$ is the eigen vector of the operator ${\cal L}_0$, 
\begin{align}
	{\cal L}_0 {\bm v}g_{\rm eq}=-\frac{1}{\tau_{\rm D}}{\bm v}g_{\rm eq}, 
	\label{eq;eigenL0}
\end{align}
and ${\cal P} {\bm v}g_{\rm eq}=0$;
we therefore obtain\begin{align}
	\exp\left[(t-t_1) {\cal L}_0 \right] {\cal Q}
	{\cal L}_{\rm L}  g_{\rm eq} \rho (\bm{r}, t_1) =
	\exp\left[-(t-t_1)/\tau_{\rm D} \right] 
	{\cal L}_{\rm L}  g_{\rm eq} \rho (\bm{r}, t_1) .
	\label{eq:projectionp1}
\end{align}
Part of Eq. (\ref{eq:formalsolP3}) can be expressed as 
\begin{align}
	-\int \mbox{d} \bm{v} 
	\bm{v} \cdot \frac{\partial}{\partial \bm{r}} \exp\left[(t-t_1) {\cal L}_0 \right] {\cal Q}
	{\cal L}_{\rm L}  g_{\rm eq} \rho (\bm{r}, t_1) =-\int \mbox{d} \bm{v} 
	 \bm{v} \cdot \frac{\partial}{\partial \bm{r}} \exp\left[-(t-t_1)/\tau_{\rm D} \right] 
	{\cal L}_{\rm L}  g_{\rm eq} \rho(\bm{r}, t_1).
	\label{eq;projectpart1}
\end{align}
Using Eq. (\ref{eq:2_13}), we obtain 
\begin{align}
	-\int \mbox{d} \bm{v} 
	\bm{v} \cdot \frac{\partial}{\partial \bm{r}} & \int_0^t dt_1\exp\left[(t-t_1) {\cal L}_0 \right] {\cal Q}
	{\cal L}_{\rm L}  g_{\rm eq}  \rho(\bm{r},t_1)  = 
	\nonumber\\
	& \int_0^t dt_1\, \exp\left[-(t-t_1)/\tau_{\rm D} \right]
	\frac{\partial}{\partial \bm{r}} \cdot 
	\left[ \frac{D}{\tau_{\rm D}}\left(\frac{\partial}{\partial \bm{r}} + \frac{1}{k_{\rm B} T} \frac{\partial U}{\partial \bm{r}} \right) 
	\right]  \rho(\bm{r},t_1). 
	\label{eq;projectpart2}
\end{align}

Similarly, we consider 
\begin{align}
	-\int \mbox{d} \bm{v} 
	\bm{v} \cdot \frac{\partial}{\partial \bm{r}} & \int_0^t dt_1\exp\left[(t-t_1) {\cal L}_0 \right] {\cal Q}
	{\cal L}_{\rm R}  g_{\rm eq}  \rho(\bm{r},t_1) . 
	\label{eq;projectpartvLR}
\end{align}
By introducing Eq. (\ref{eq:gdv}), we have
\begin{align}
	{\cal L}_{\rm R} g_{\rm eq}\rho(\bm{r}, t_1)&=-\bm{n} \cdot 
	\bm{v}f_{\rm r} (|v_z|)\theta(-\bm{n} \cdot \bm{v}) \delta(r-\sigma) g_{\rm eq}\rho(\bm{r}, t_1) .
	\label{eq:LRgeq}
\end{align}
Using Eq. (\ref{eq;eigenL0}) and ${\cal P} {\bm v}g_{\rm eq}=0$, we obtain 
\begin{align}
	\int_0^t dt_1\exp\left[(t-t_1) {\cal L}_0 \right] {\cal Q}
	{\cal L}_{\rm R}  g_{\rm eq} \rho(\bm{r}, t_1)  =\int_0^t dt_1 \exp\left[-(t-t_1)/\tau_{\rm D} \right] 
	{\cal L}_{\rm R}  g_{\rm eq} \rho(\bm{r}, t_1) .
	\label{eq;projectpart3}
\end{align}
By introducing Eq. (\ref{eq:24}), we find 
\begin{align}
	\int d \bm{v} \bm{v} {\cal L}_{\rm R}  g_{\rm eq}  \rho(\bm{r},t_1)=\vec{\bm{\kappa}} \rho(\bm{r}, t_1) \frac{\delta(r-\sigma)}{4\pi \sigma^2}
	\label{eq;projectpart4}
\end{align}
and obtain
\begin{align}
	\int \mbox{d} \bm{v} 
	\bm{v} \cdot \frac{\partial}{\partial \bm{r}} & \int_0^t dt_1\exp\left[(t-t_1) {\cal L}_0 \right] {\cal Q}
	{\cal L}_{\rm R}  g_{\rm eq}  \rho(\bm{r},t_1)  =
	\nonumber \\
	& \int_0^t dt_1\, \exp\left[-(t-t_1)/\tau_{\rm D} \right] \frac{\partial}{\partial \bm{r}} \cdot  \vec{\bm{\kappa}} \rho(\bm{r}, t_1) \frac{\delta(r-\sigma)}{4\pi \sigma^2} .
	\label{eq;projectpart5}
\end{align}

Finally, we consider 
\begin{align}
	\int \mbox{d} \bm{v} 
	{\cal L}_{\rm R}  \int_0^t dt_1& \exp\left[(t-t_1) {\cal L}_0 \right] {\cal Q}
	{\cal L}_1 g_{\rm eq} \rho (\bm{r}, t_1) \approx 
	\nonumber \\
	&
	\int \mbox{d} \bm{v} 
	{\cal L}_{\rm R}  \int_0^t dt_1\exp\left[(t-t_1) {\cal L}_0 \right] {\cal Q}
	{\cal L}_{\rm L} g_{\rm eq} \rho(\bm{r}, t_1), 
	\label{eq:formalsolP6}
\end{align}
in Eq. (\ref{eq:formalsolP3}), where we substituted ${\cal L}_{\rm L}$ for ${\cal L}_1$ as the lowest order in the perturbation expansion. 
By substituting Eq. (\ref{eq:LLgeq}) and using Eq. (\ref{eq:projectionp1}), we obtain 
\begin{align}
	\int \mbox{d} \bm{v} 
	&{\cal L}_{\rm R}  \int_0^t dt_1\exp\left[(t-t_1) {\cal L}_0 \right] {\cal Q}
	{\cal L}_{\rm L} g_{\rm eq} \rho (\bm{r}, t_1)  = \nonumber \\
	&-\int \mbox{d} \bm{v} 
	{\cal L}_{\rm R}  \int_0^t dt_1\exp\left[-(t-t_1)/\tau_{\rm D} \right] 
	{\bm v}\cdot \left(\frac{\partial}{\partial \bm{r}} + \frac{1}{k_{\rm B} T} \frac{\partial U}{\partial \bm{r}} \right) g_{\rm eq}\rho (\bm{r}, t_1) .
	\label{eq:formalsolP7}
\end{align}
Using Eq. (\ref{eq;projectpart4}), we can express Eq. (\ref{eq:formalsolP7}) as 
\begin{align}
	\int \mbox{d} \bm{v} 
	&{\cal L}_{\rm R}  \int_0^t dt_1\exp\left[(t-t_1) {\cal L}_0 \right] {\cal Q}
	{\cal L}_{\rm L} g_{\rm eq} \rho (\bm{r}, t_1) = \nonumber \\
	&-\int_0^t dt_1\exp\left[-(t-t_1)/\tau_{\rm D} \right] 
	\frac{\delta(r-\sigma)}{4\pi \sigma^2}\vec{\bm{\kappa}}\cdot \left(\frac{\partial}{\partial \bm{r}} + \frac{1}{k_{\rm B} T} \frac{\partial U}{\partial \bm{r}} \right) \rho(\bm{r}, t_1)  \nonumber \\
	&=0, \mbox{ for } \epsilon \rightarrow 0, 
	\label{eq:formalsolP8}
\end{align}
where the reflecting boundary condition given by Eq. (\ref{eq:bc_2_26}) is introduced in the limit of $\epsilon \rightarrow 0$. 
In principle, the limit of $\epsilon \rightarrow 0$ should be taken after obtaining the reaction rate coefficient; however, this term can be shown to not contribute to the final expression.
Note that the derivative with respect to ${\bm r}$ is not applied to the delta-function.  
If the derivative with respect to ${\bm r}$ is applied to the delta-function, as in Eq. (\ref{eq;projectpart5}), we should not take this limit. 

By collecting Eqs. (\ref{eq:formalsolP3}), (\ref{eq;projectpart2}), (\ref{eq;projectpart5}), and (\ref{eq:formalsolP8}), we obtain
\begin{align}
	\frac{\partial}{\partial t} \rho(\bm{r}, t) = &
	\int_0^t dt_1 \exp\left(-\frac{t-t_1}{\tau_{\rm D}} \right)
	\frac{\partial}{\partial \bm{r}} \cdot 
	\left[ \frac{D}{\tau_{\rm D}}\left(\frac{\partial}{\partial \bm{r}} + \frac{1}{k_{\rm B} T} \frac{\partial U}{\partial \bm{r}} \right) 
	\right]  \rho(\bm{r},t_1)-\kappa  \rho(\bm{r}, t) \frac{\delta(r-\sigma)}{4\pi \sigma^2}
	\nonumber \\
	& -
	\int_0^t dt_1\exp\left(-\frac{t-t_1}{\tau_{\rm D}} \right) 
	\frac{\partial}{\partial \bm{r}} \cdot \vec{\bm{\kappa}} \rho(\bm{r}, t_1) \frac{\delta(r-\sigma)}{4\pi \sigma^2} , 
	\label{eq:formalsolP9}
\end{align}
where $\kappa$ is given by Eq. (\ref{eq:28}) and $\vec{\bm{\kappa}}$ can be expressed as $\vec{\bm{\kappa}}=\kappa_{\rm r} \bm{n}$ using $\kappa_{\rm r}$ given by Eq. (\ref{eq:30}). 
In this way, Eq. (\ref{eq:2_31_tm}) is reproduced using the projection operator method. 
When the time convolution is decoupled, the time integration gives the factor $\tau_{\rm D}$ multiplied to $D/\tau_{\rm D}$ and $\vec{\bm{\kappa}}$ in Eq. (\ref{eq:formalsolP9}). 
In isotropic systems, we can express $\tau_{\rm D} \kappa_{\rm r}= 2 \pi \sigma^2 D$ using Eqs. (\ref{eq:2_13}) and (\ref{eq:30}).

Thus far, we have considered the case of collision-induced reactions, where the intrinsic reaction rate constant depends on the velocity, as shown by the multiplication factor of $f(\sigma, \bm{v},t)$ in Eq. (\ref{eq:bc1_1}). 
For electron transfer and energy transfer, the intrinsic reaction rate constant can be independent of the reactant velocity. 
When the intrinsic reaction rate constant is independent of the reactant velocity and is localized, the intrinsic reaction rate constant can be expressed as
\begin{align}
	R(\bm{v},\bm{r},t)=\frac{\kappa_{\rm i}}{4\pi \sigma^2}f(\sigma, \bm{v},t) \delta(r-\sigma)  .
	\label{eq:32}
\end{align}
We can express ${\cal L}_{\rm R}$ as 
\begin{align}
	{\cal L}_{\rm R} &=-\kappa_{\rm i}f(\sigma, \bm{v},t) \delta(r-\sigma) 
	\label{eq:LRi}
\end{align}
instead of Eq. (\ref{eq:LR}) and obtain
$\int \mbox{d} \bm{v}  {\cal L}_{\rm R} g_{\rm eq}=-[\kappa_{\rm i}/(4\pi \sigma^2) ]\delta(r-\sigma)$. 
We then have
\begin{align}
	\int d \bm{v} \bm{v} {\cal L}_{\rm R}  g_{\rm eq}  \rho(\bm{r},t_1)=0 ,
	\label{eq;projectpart41}
\end{align}
instead of Eq. (\ref{eq;projectpart4}) because $\vec{\bm{\kappa}}=0$ when the intrinsic reaction rate constant is given by Eq. (\ref{eq:32}). 
In this case, we obtain
\begin{align}
	\frac{\partial}{\partial t} \rho(\bm{r}, t) &= 
	\int_0^t dt_1 \exp\left(-\frac{t-t_1}{\tau_{\rm D}} \right)
	\frac{\partial}{\partial \bm{r}} \cdot 
	\left[ \frac{D}{\tau_{\rm D}}\left(\frac{\partial}{\partial \bm{r}} + \frac{1}{k_{\rm B} T} \frac{\partial U}{\partial \bm{r}} \right) 
	\right]  \rho(\bm{r},t_1) -\kappa_{\rm i}  \rho(\bm{r}, t) \frac{\delta(r-\sigma)}{4\pi \sigma^2} 
	\label{eq:formalsolP10}
\end{align}
instead of Eq. (\ref{eq:formalsolP9}). 

%%%%%%%%%%%%%%%%%%%%%%%%%%%%%%%%%%%%%%%%%%%%%%%%%%%%%%%%%%%%%%%%%%%%%%%%%%%%%%%%%%%%%%%%
\section{Analytical solution}
\label{sec:V}

We solve Eq. (\ref{eq:formalsolP9}) in the Laplace domain for $U=0$ expressed as 
\begin{align}
	s\hat{\rho}(r,s)-
	\rho(r,0)
	&=\frac{1}{r^2}\frac{\partial}{\partial r} D(s) r^2 
	\frac{\partial}{\partial r}
	\hat{\rho}(r,s)
	\nonumber \\
	&
	-\kappa \hat{\rho}(r,s)
	\frac{\delta(r-\sigma)}{4\pi r^2}-
	\tau_{\rm D}(s)  \frac{1}{r^2}\frac{\partial}{\partial r} r^2 \kappa_{\rm r}\hat{\rho}(r,s) \frac{\delta(r-\sigma)}{4\pi r^2} , 
	\label{eq:2_31_P}
\end{align}
where $D(s)=D/(1+s\tau_{\rm D})$ and $\tau_{\rm D}(s)=\tau_{\rm D}/(1+s\tau_{\rm D})$.  
The long-time limit of the rate coefficient as well as the initial rate coefficient will be derived from the analytical expression of the rate coefficient in the Laplace domain. 

The Green's function for
\begin{align}
	s\hat{\rho}(r,s)-\frac{\delta(r-r_{\rm i})}{4\pi r^2}
	&=\frac{1}{r^2}\frac{\partial}{\partial r} r^2 D(s) 
	\frac{\partial}{\partial r}
	\hat{\rho}(r,s) ,
	\label{eq:green}
\end{align}
is given by 
\begin{align}
	\hat{G}_\epsilon \left( r, r_{\rm i}, s
	\right) =
	\frac{1}{8\pi D(s)\zeta(s)r r_{\rm i}}
	\left[\exp \left[ - |r-r_{\rm i}| \zeta(s) \right] - 
	\frac{1-\sigma_{\rm i}\zeta(s)}{1+\sigma_{\rm i} \zeta(s)} 
	\exp \left[ - (r+r_{\rm i}-2\sigma_{\rm i}) \zeta(s) \right] 
	\right] , 
	\label{eq:8_16}
\end{align}
where  
$\zeta(s)=\sqrt{s/D(s)}$,  and 
$\sigma_{\rm i}=\sigma -\epsilon$ is introduced because of the reflecting boundary condition at $r= \sigma -\epsilon$. 
The formal solution can be expressed using the Green function as
\begin{align}
\hat{\rho}(r,s)=&\frac{1}{s}-\kappa \hat{G}_\epsilon (r,\sigma,s)  \hat{\rho}(\sigma,s) -
\nonumber \\
&4\pi\tau_{\rm D}(s) \int_{\sigma-\epsilon}^\infty d r_0  \hat{G}_\epsilon (r,r_0,s)   \frac{\partial}{\partial r_0} r_0^2\kappa_{\rm r}\hat{\rho}(r_0,s)\frac{\delta(r_0-\sigma)}{4\pi r_0^2} , 
\label{eq:rhors_1}
\end{align}
where we used
\begin{align}
	\lim_{\epsilon \rightarrow 0}\int_{\sigma-\epsilon}^\infty 4\pi r_{\rm i}^2 dr_{\rm i} \hat{G}_\epsilon(\sigma,r_{\rm i},s) \rho(r_{\rm i},0)=1  , 
	\label{eq:2_37}
\end{align}
for the initial uniform distribution with a reflecting boundary condition at $r=\sigma-\epsilon$.
By performing the partial integration, the last term in Eq. (\ref{eq:rhors_1}) can be rewritten as
\begin{align}
4\pi\tau_{\rm D}(s) \int_{\sigma-\epsilon}^\infty d r_0  \hat{G}_\epsilon (r,r_0,s)  & \frac{\partial}{\partial r_0} r_0^2\kappa_{\rm r}\hat{\rho}(r_0,s)\frac{\delta(r_0-\sigma)}{4\pi r_0^2}
\nonumber \\
&
=-\tau_{\rm D}(s)\left. \kappa_{\rm r}\hat{\rho}(\sigma,s) \frac{\partial}{\partial r_0} \hat{G}_\epsilon (r,r_0,s) \right|_{r_0=\sigma}. 
\label{eq:rhors_3}
\end{align}
By substituting Eq. (\ref{eq:rhors_3}) into Eq. (\ref{eq:rhors_1}), a closed equation for $\hat{\rho}(\sigma,s)$ can be obtained,   
\begin{align}
\hat{\rho}(\sigma,s)=&\frac{1}{s}-\kappa \hat{G}_\epsilon (\sigma,\sigma,s)  \hat{\rho}(\sigma,s) +
\tau_{\rm D}(s)\left. \kappa_{\rm r}\hat{\rho}(\sigma,s) \frac{\partial}{\partial r_0} \hat{G}_\epsilon (\sigma,r_0,s) \right|_{r_0=\sigma} .
\label{eq:rhors_4}
\end{align}
There remains a subtlety in the boundary condition.
Depending on the limit, we have
\begin{align}
	\lim_{\epsilon \rightarrow 0}\lim_{r \rightarrow \sigma} \frac{\partial}{\partial r}\hat{G}_\epsilon \left( \sigma, r, s
	\right) &=0 \mbox{ for } \sigma>r ,
	\label{eq:Grgri}\\
	\lim_{\epsilon \rightarrow 0}\lim_{r \rightarrow \sigma} \frac{\partial}{\partial r}\hat{G}_\epsilon \left( \sigma, r, s
	\right) &=-\frac{1}{4\pi \sigma^2 D(s)} \mbox{ for } r>\sigma  .
	\label{eq:Grigr}
\end{align}
$\hat{G}_\epsilon \left( \sigma, r, s
	\right)$ is not differentiable at the point $r=\sigma$
where the first derivative of $|x|$ with respect to $x$ is used when 
performing the derivative with respect to $r_i$ for Eq. (\ref{eq:8_16}) by applying the chain rule for calculating the derivative of composition of $|x|$ and 
$x=r-r_{\rm i}$. 
We, therefore, consider Eq. (\ref{eq:Grgri}) or (\ref{eq:Grigr}). 
In the limit of $\epsilon \rightarrow 0$,  
$r$ approaches $\sigma$ from the side satisfying $r>\sigma$ rather than the side satisfying $r<\sigma$;  
by considering that the physical meaningful limit towards the reaction surface is $ r \rightarrow \sigma+$,  
we use Eq. (\ref{eq:Grigr}) rather than Eq. (\ref{eq:Grgri}). 
Equation (\ref{eq:Grigr}) is also consistent with the previous theoretical model, where a division into inner and outer spatial regions separated by a boundary is introduced. \cite{Northrup_78}
By adopting Eq. (\ref{eq:Grigr}) and using Eq. (\ref{eq:rhors_4}), we obtain
\begin{align}
	\hat{\rho}(\sigma,s)
	&=\lim_{\epsilon \rightarrow 0}\frac{1}{s}\left(1 +\kappa \hat{G}_\epsilon(\sigma,\sigma,s) +\frac{\tau_{\rm D} }{4\pi \sigma^2 D} \kappa_{\rm r}
	\right)^{-1} , 
	\label{eq:2_35}
\end{align}

Using Eq. (\ref{eq:2_32}), we obtain the Laplace transform of the reaction rate coefficient, $k(t)$, as 
\begin{align}
	\hat{k}(s)=\lim_{\epsilon \rightarrow 0}\kappa \hat{\rho}(\sigma,s)=
	\frac{\kappa}{s}\left(1 +\kappa \hat{G}_0(\sigma,\sigma,s) +\frac{\tau_{\rm D} }{4\pi \sigma^2 D} \kappa_{\rm r}
	\right)^{-1} ,
	\label{eq:2_38}
\end{align}
where the limit of $\epsilon \rightarrow 0$ is taken %because the functional form will not be changed by shifting $\epsilon$ after the inverse Laplace transformation; 
and $\hat{G}_0(\sigma,\sigma,s)$ can be expressed as 
\begin{align}
	\hat{G}_0(\sigma,\sigma,s) &= \frac{1+s\tau_{\rm D}}{4\pi \sigma D} \frac{1}{1+\sigma \sqrt{s(1+s\tau_{\rm D})/D}} . 
	\label{eq:2_36_1}
\end{align}
Assuming that in Eq. (\ref{eq:2_35}) the limits as $r_0, r \to \sigma$ 
and inverse Laplace transformation are commutative, we can conclude: 
Equation (\ref{eq:2_38}) is the exact reaction rate coefficient in the Laplace domain for the Cattaneo--Vernotte differential model.

When $\kappa_{\rm r}=0$, Eq. (\ref{eq:2_38}) can be expressed as
\begin{align}
	\hat{k}(s)=\lim_{\epsilon \rightarrow 0}\kappa \hat{\rho}(\sigma,s)=\frac{1}{s}\left(\frac{1}{\kappa} +\hat{G}_0(\sigma,\sigma,s) 
	\right)^{-1} .
	\label{eq:KK}
\end{align}

The long-time reaction rate coefficient can be obtained as 
\begin{align}
	k_\infty &=  \lim_{\epsilon \rightarrow 0}\lim_{s \rightarrow 0}
	\left(\frac{1}{\kappa} +\hat{G}_\epsilon(\sigma,\sigma,s) \right)^{-1} \\
	&=\left[\frac{1}{\kappa} \left(1  +\frac{\tau_{\rm D}}{\sigma} \frac{\kappa_{\rm r}}{4\pi \sigma D} \right) +\frac{1}{4\pi \sigma D} 
	\right]^{-1} 
	\label{eq:ks1}
	\\
	&=\left[\frac{1}{\kappa} \left(1+\frac{f_{\rm r}}{2}
	\right) +\frac{1}{4\pi \sigma D}
	\right]^{-1} ,
	\label{eq:ks2}
\end{align}
where Eq. (\ref{eq:30}) is substituted and $\kappa$ is given by Eq. (\ref{eq:28}). (The result is the same as that obtained from $k_\infty = \lim_{s \rightarrow 0} s\hat{k}(s)$.) 
For collision-induced reactions without reflection at the contact distance, we have $f_{\rm r}=1$. 
For $f_{\rm r}=1$, $\kappa$ is effectively reduced to $2\kappa/3$ in Eq. (\ref{eq:ks2}). 
As explained below Eq. (\ref{eq:Rst}), the inward current density is reduced by the positive correlation between the velocity vector and the inward normal component of the velocity vector at the contact distance.  
This effect reduces $\kappa$ to $2\kappa/3$ for collision-induced reactions without reflection at the contact distance in the lowest order of the perturbation expansion for the reaction sink term and the streaming term. 
The persistence of inertial effects on the reduction of the reaction rate coefficient has been also observed by Langevin dynamic simulations. \cite{Yang_01} 

The aforementioned results can be compared to the long-time rate coefficient obtained by assuming $\kappa_{\rm r}=0$ given by 
\begin{align}
	k_\infty^{(0)} &= \left[\frac{1}{\kappa} +\frac{1}{4\pi \sigma D} 
	\right]^{-1} .
	\label{eq:ksconv}
\end{align}

In the opposite limit of $t\rightarrow 0$, $k_{\rm f} =k(0)$ can be estimated from $k_{\rm f}=\lim_{s \rightarrow \infty} s k(s)$. 
We obtain
\begin{align}
	k_{\rm f} =\left[\frac{1}{\kappa} \left(1  +\frac{\tau_{\rm D}}{\sigma} \frac{\kappa_{\rm r}}{4\pi \sigma D} \right) +\frac{1}{4\pi \sigma^2} \sqrt{\frac{\tau_{\rm D}}{D}}
	\right]^{-1} 
	\label{eq:kf1}
\end{align}
using Eq. (\ref{eq:2_38})
and obtain
\begin{align}
	k_{\rm f}^{(0)} &= \left[\frac{1}{\kappa} +\frac{1}{4\pi \sigma^2} \sqrt{\frac{\tau_{\rm D}}{D}}
	\right]^{-1} 
	\label{eq:kfconv}
\end{align}
by assuming $\kappa_{\rm r}=0$.
The physical origin of this sudden drop of the initial rate coefficient from the equilibrium value $\kappa$ was discussed in Ref. \onlinecite{Lee_23}.

%%%%%%%%%%%%%%%%%%%%%%%%%%%%%%%%%%%%%%%%%%%%%%%%%%%%%%%%%%%%%%%%%%
\section{Numerical results}
\label{se:VI}

The Fokker--Planck--Kramers equation is the continuous description in phase space under Markovian momentum relaxation. \cite{KRAMERS_40,DOI_75,Northrup_78,Naqvi_82,Harris_83,MOLSKI_88,Ibuki_97,Ibuki_03,Ibuki_06,Kim_09}
The assumption of Markovian momentum relaxation might be justified if the mean free path is sufficiently smaller than the contact distance. 
$\sqrt{k_{\rm B} T/(2\mu)}$ indicates the mean velocity of reactants in thermal equilibrium, and $\ell_v=\sqrt{k_{\rm B} T/(2\mu)}\, \tau_{\rm D}$ can be interpreted as the mean free path that reactants are able to travel during the momentum relaxation time ($\tau_{\rm D}$) when a reaction is absent.
The condition that the mean free path is smaller than the contact distance can be expressed by 
\begin{align}
	\tau_{\rm D} \sqrt{\frac{k_{\rm B} T}{2\mu}} <\sigma  . 
	\label{eq:2_40}
\end{align}
If Eq. (\ref{eq:2_40}) is satisfied, momentum can be relaxed by successive collision-induced events and the memory of successive collision-induced events can be lost before the reactant moves at a distance characterized by $\sigma$. 
The mean free path, $\ell_v$, divided by the contact distance corresponds to the Knudsen number in hydrodynamics, which is required for a continuum description of reactants. 
Under the condition $\ell_v/\sigma <1$, we take into account the effect of a reaction in the Fokker--Planck--Kramers equation.  
In this section, we present the results under $\ell_v/\sigma <1$.

For simulations, we rewrite Eqs. (\ref{eq:2_22})--(\ref{eq:2_23}) with $U=0$. 
By multiplying both sides of Eq. (\ref{eq:2_22}) by $4\pi \int_{\sigma-\epsilon}^{\sigma+\epsilon} dr r^2$ and taking the limit of $\epsilon \rightarrow 0$, we obtain
\begin{align}
	\frac{\partial}{\partial t} \rho(r,t)&+
	\frac{1}{r^2}\frac{\partial}{\partial r} r^2 j_{r}(r,t)=0, 
	\label{eq:2_26}
\end{align} 
with the boundary condition given by 
\begin{align}
	j_{\rm r}(\sigma,t)&= 
	-\frac{\kappa \rho(\sigma,t)}{4\pi \sigma^2} ,
	\label{eq:bc_2_18_1}
\end{align}
which is equivalent to Eq. (\ref{eq:2_22}) with the boundary condition given by $j_{\rm r}(\sigma-\epsilon,t)=0$ if the limit of $\epsilon \rightarrow 0$ limit is taken when calculating the reaction rate coefficient. 
The kinetic equation for $ j_{\rm r} (r,t)$ is given by Eq. (\ref{eq:2_23}) with $U=0$ as
\begin{align}
	\frac{\partial}{\partial t} j_{\rm r}(r,t)&=-\frac{1}{\tau_{\rm D}} 
	\left[ j_{\rm r} (r,t)+
	D\frac{\partial}{\partial r} \rho(r,t)-\tau_{\rm D}\kappa_{\rm r} \rho(r,t)
	\frac{\delta(r-\sigma)}{4\pi r^2}\right] . 
	\label{eq:2_23_f2}
\end{align}

We here study the effect of the momentum relaxation time ($\tau_{\rm D}$) on the long-time rate coefficient. 
For this purpose, we use $\tau_{\rm c}=\sigma \sqrt{2\mu/(k_{\rm B} T)}$ as the time unit. 
We introduce dimensionless radial coordinate $\zeta=r/\sigma$, time $\tau=t/\tau_{\rm c}$, $J_{\rm r}=j_{\rm r} \tau_{\rm c}/\sigma$, $D_{\rm n}= D\tau_{\rm c}/\sigma^2 =2\tau_{\rm D}/\tau_{\rm c}$, $\kappa_{\rm n}=\kappa\tau_{\rm c}/\sigma^3  =4\sqrt{\pi} f_{\rm r}$, and $\kappa_{\rm rn}=\kappa_{\rm r} \tau_{\rm D} \tau_{\rm c}/\sigma^4=2 \pi( \tau_{\rm D}/\sigma)\sqrt{2k_{\rm B} T/\mu}\, f_{\rm r}$, which is proportional to $\tau_{\rm D}$.
By considering that $D=\tau_{\rm D} k_{\rm B} T/\mu$ is also proportional to $\tau_{\rm D}$, we express $\kappa_{\rm rn}=2 \pi D_{\rm n} f_{\rm r}$ for numerical evaluation. 

%%%%%%%%%%%%%%%%%%%%%%%%%%%%%%%%%%%%%%%%%%%%%%%%%%%%%%%%%%%%%%$\bm{\kappa}_{\rm hG}=2\bm{\kappa}$
\begin{figure}[h]
	\includegraphics[width=\textwidth]{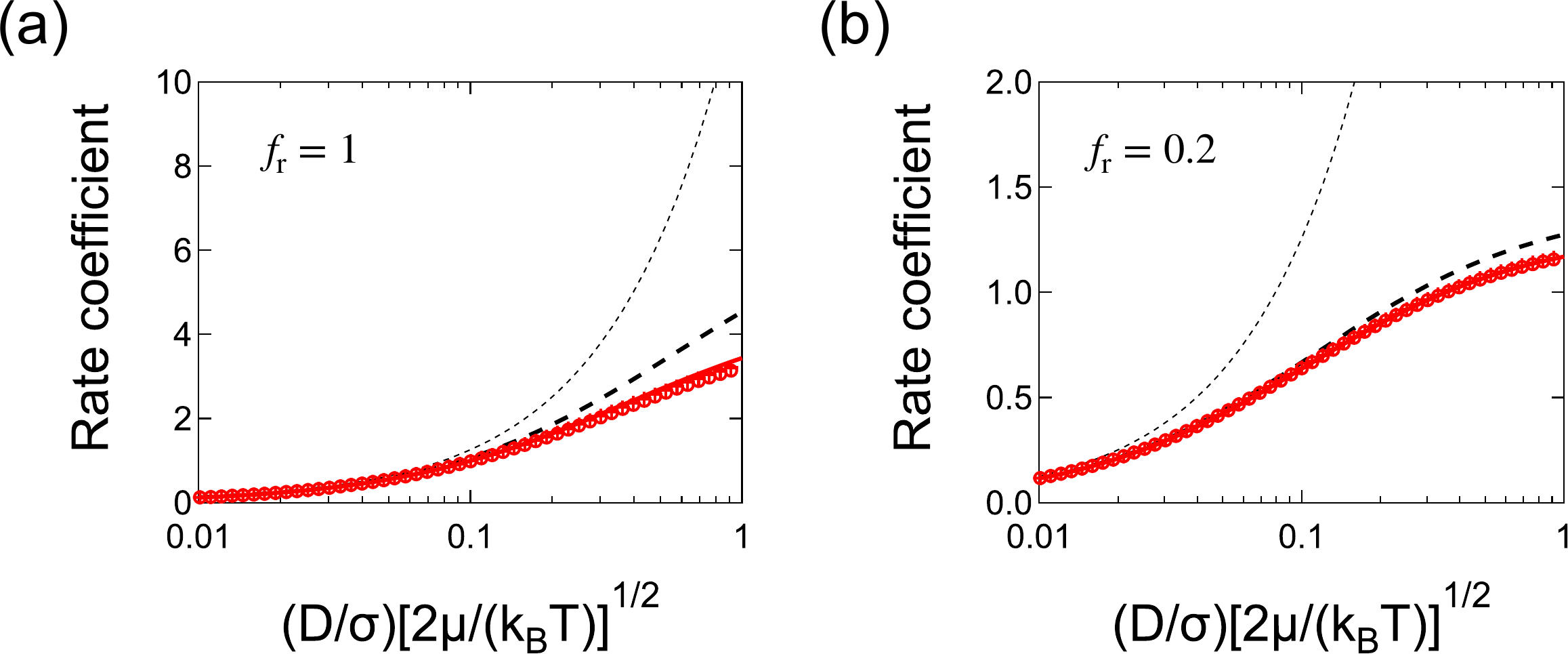}
	%\vspace{1.5cm}
	\caption{(Color online) 
		Dimensionless rate coefficient [$(k_\infty/\sigma^2)\sqrt{2\mu/(k_{\rm B} T)}$] in the long-time limit obtained using Eqs. (\ref{eq:r_4})--(\ref{eq:r_6}) is shown against $D_{\rm n}=(D/\sigma)\sqrt{2\mu/(k_{\rm B} T)}$ for $(\kappa/\sigma^2)\sqrt{2\mu/(k_{\rm B} T)}=4\sqrt{\pi} f_{\rm r}$, where $\zeta_{\rm max}=100$.   
		(a) and (b) show the results for $f_{\rm r}=1$ and $f_{\rm r}=0.2$, respectively.
		Circles and crosses indicate the numerical results obtained using the Lorentzian representation of the delta-function $2\eta/[\pi(\zeta^2+\eta^2)]$ with $\eta =10^{-6}$ and the Lorentzian representation of the delta-function $\eta/[\pi(\zeta^2+\eta^2)]$ located at distance $5\eta$ shifted outward from the reflecting boundary, respectively.
		The black short-dashed line is the diffusion-controlled limit of $4\pi D_{\rm n}$.
		The black long-dashed line indicates the results for $\kappa_{\rm r}=0$ given by Eq. (\ref{eq:ksconv}). 
		The red thick line indicates the results obtained using Eq. (\ref{eq:ks2}). 
	}
	\label{fig:1}
\end{figure}

Using Eqs. (\ref{eq:2_22}) and (\ref{eq:2_23_f2}), we obtain the rate coefficient in the long-time limit from $k_{\infty\rm{n}}=4\pi J_{\rm rs} (1)$ by numerically solving 
\begin{align}
	\frac{\partial}{\partial \zeta} \zeta^2 J_{\rm rs}(\zeta)&=0, 
	\label{eq:r_4}\\
	J_{\rm rs} (\zeta)+
	D_{\rm n}\frac{\partial}{\partial \zeta} \rho_{\rm s}(\zeta)
	-\kappa_{\rm rn} \rho_{\rm s}(\zeta)
	\frac{\delta(\zeta-1)}{4\pi}&=0 , 
	\label{eq:r_5}
\end{align} 
with the boundary condition given by
\begin{align}
	J_{\rm rs} (1)=-\frac{\kappa_{\rm n}}{4\pi} \rho_{\rm s}(1) ,
	\label{eq:r_6}
\end{align}
and $\rho_{\rm s}(\zeta_{\rm max})=1$; 
steady states are labeled with the subscript ``s''. 
We use the Lorentzian representation of the delta-function given by $f_{\rm d1}(\zeta)=2\eta/[\pi(\zeta^2+\eta^2)]$ with $\eta =10^{-6}$, where the normalization is given by $\int_0^\infty dx f_{\rm d1}(\zeta)=1$. 
We also use the Lorentzian representation of the delta-function given by $f_{\rm d2}(\zeta)=\eta/[\pi(\zeta^2+\eta^2)]$ by changing the delta-function in Eq. (\ref{eq:r_5}) by $\delta(\zeta-1-5\eta)$  by noticing $\int_1^\infty d\zeta f_{\rm d2}(\zeta)\approx 0.94$. 
The factor $5 \eta$ with $\eta =10^{-6}$ corresponds to $\epsilon$ 
representing the difference between the contact distance  ($\sigma$), where the reaction sink term is located,  
and $\sigma -\epsilon$, where the reflecting boundary is located;  
here, the reflecting boundary is set at $\zeta=1$.
As shown in Fig. \ref{fig:1}, the numerical results for $k_{\infty\rm{n}}$ obtained using Eqs. (\ref{eq:r_4})--(\ref{eq:r_6}) and $\zeta_{\rm max}=100$ are not influenced by the choice of Lorentzian representation nor by changing $\eta$ to $\eta =10^{-4}$ (results not shown) for $f_{\rm r}=1$ and $f_{\rm r}=0.2$.  
The results of Eq. (\ref{eq:ksconv}), where $\kappa_{\rm r}$ is ignored, can be regarded as the upper bound of the long-time rate coefficient. 
The numerical results are close to 
the exact results obtained from Eq. (\ref{eq:ks2}); they are consistent with each other.

By decreasing $D$, which is proportional to $\tau_{\rm D}$, we can approximate the long-time rate coefficient by the diffusion-controlled rate. 
In Fig.  \ref{fig:1}, the diffusion-controlled limit is shown by the black short-dashed line. 
The diffusion-controlled rate coefficient is given by $4 \pi \sigma D$, which is proportional to the contact distance. 
In the reaction-controlled limit, the long-time rate coefficient is limited by the rate constant expressed by $\kappa=2\sigma^2\sqrt{2\pi k_{\rm B} T/\mu}\, f_{\rm r}$ [Eq. (\ref{eq:28})], which is proportional to the square of the contact distance because of the scattering cross-sectional area for the ballistic motion of reactants. 
With an increase in $D$, the long-time rate coefficient is more limited by the reaction. 

%%%%%%%%%%%%%%%%%%%%%%%%%%%%%%%%%%%%%%%%%%%%%%%%%%%%%%%%%%%%%%$\bm{\kappa}_{\rm hG}=2\bm{\kappa}$
\begin{figure}[h]
	\includegraphics[width=0.5\textwidth]{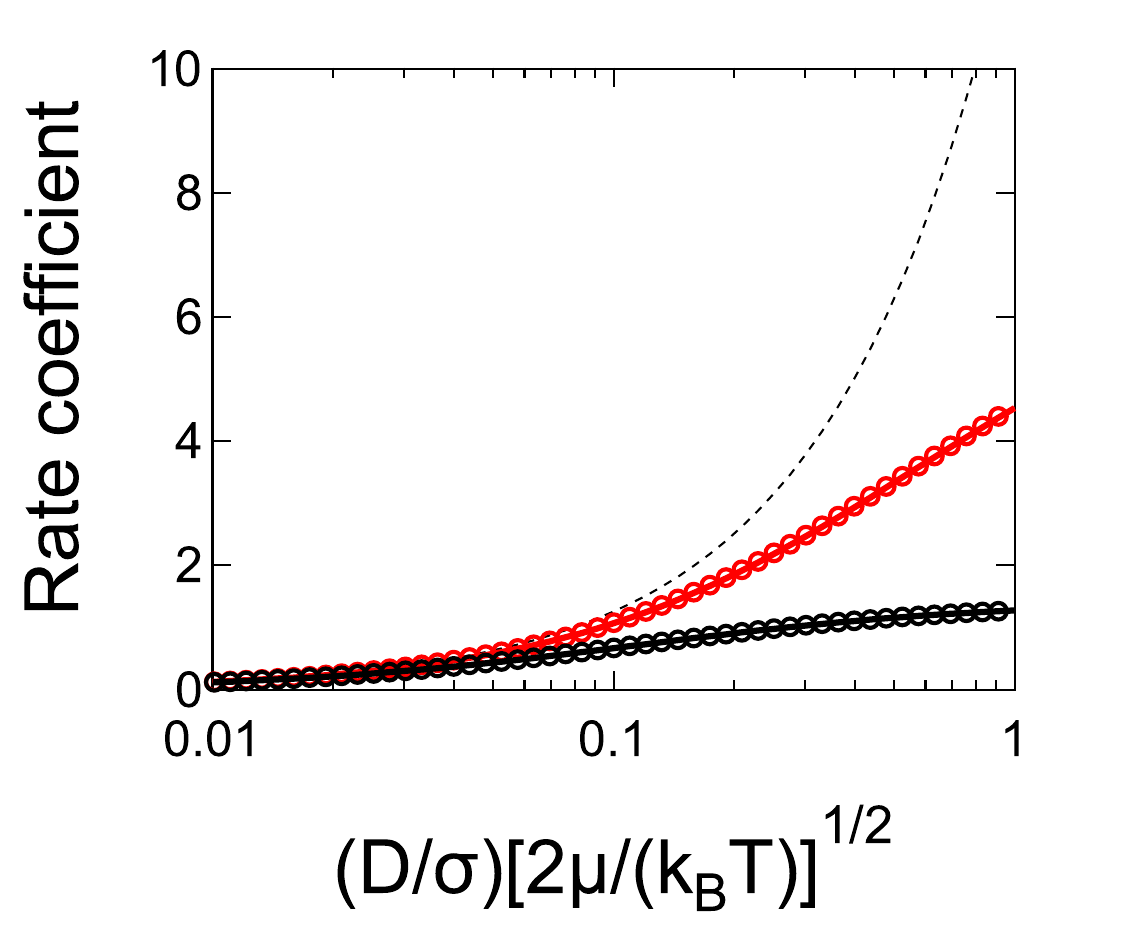}
	%\vspace{1.5cm}
	\caption{(Color online) 
		The dimensionless rate coefficient [$(k_\infty/\sigma^2)\sqrt{2\mu/(k_{\rm B} T)}$] in the long-time limit obtained using Eqs. (\ref{eq:r_4})-(\ref{eq:r_6}) for $\kappa_{\rm r}=0$ is shown against $D_{\rm n}=(D/\sigma)\sqrt{2\mu/(k_{\rm B} T)}$ for $(\kappa/\sigma^2)\sqrt{2\mu/(k_{\rm B} T)}=4\sqrt{\pi} f_{\rm r}$, where $\zeta_{\rm max}=100$. 
		The upper red solid lines and circles and the lower black solid lines and circles indicate $f_{\rm r}=1$ and $f_{\rm r}=0.2$, respectively.
		Circles indicate the numerical solutions of Eqs. (\ref{eq:r_4})--(\ref{eq:r_6}) for $\kappa_{\rm r}=0$. 
		The black short-dashed line is the diffusion-controlled limit of $4\pi D_{\rm n}$.
		The thick lines indicate the results for $\kappa_{\rm r}=0$ given by Eq. (\ref{eq:ksconv}). 
	}
	\label{fig:2}
\end{figure}
When the intrinsic reaction rate constant is independent of the reactant velocity and is localized, we obtain $\kappa_{\rm r}=0$, as shown by Eq. (\ref{eq;projectpart41}). 
In this case, the long-time rate constants are not influenced by the inertial effect and can be obtained from Eq. (\ref{eq:ksconv}), which is not influenced by $\tau_{\rm D}$ except through the diffusion constant. 
In Fig. \ref{fig:2}, the solid lines are the same as the black long-dashed lines in Fig. \ref{fig:1}.
The numerical solutions of Eqs. (\ref{eq:r_4})--(\ref{eq:r_6}) using $f_{\rm d1}(\zeta)$ for $\kappa_{\rm r}=0$ are close to the values obtained from Eq. (\ref{eq:ksconv}). 
Conversely, the influence of the inertial effect remains in the values of the long-time rate constants, as evidenced by the results in Fig. \ref{fig:1}, when the intrinsic reaction rate constant depends on the reactant velocity.

%%%%%%%%%%%%%%%%%%%%%%%%%%%%%%%%%%%%%%%%%%%%%%%%%%%%%%%%%%%%%%Sergey_kinetics-crop %

%%%%%%%%%%%%%%%%%%%%%%%%%%%%%%%%%%%%%%%%%%%%%%%%%%%%%%%%%%%%%%%%%%%%%
%%%%%%%%%%%%%%%%%%%%%%%%%%%%%%%%%%%%%%%%%%%%%%%%%%%%%%%%%%%%%%%%%%%%%
\section{Conclusion}
\label{sec:VII}

Although the Cattaneo--Vernotte model has been widely studied to take into account momentum relaxation in transport equations, the effect of reactions on the Cattaneo--Vernotte model has not yet been fully elucidated. 
How current density associated with reactions can be expressed in the Cattaneo--Vernotte model is unclear.
We derived the effect of a reaction on the Cattaneo--Vernotte model using the Fokker--Planck--Kramers equation. 

We took into account momentum relaxation by applying the projection operator method to the Fokker--Planck--Kramers equation with a reaction sink term describing reactions. 
In the absence of a reaction sink term, a modified Smoluchowski equation including a memory kernel could be derived. 
In the lowest order of the perturbation expansion for the reaction sink term and the streaming term, we obtain the modified Smoluchowski equation generalized to include two reaction terms [Eq. (\ref{eq:formalsolP9}) ] for collision-induced reactions, 
where the intrinsic reaction rate constant depends on the relative velocity of reactants. 
The term multiplied by $\kappa_{\rm r}$ is coupled to the memory kernel, whereas the term multiplied by $\kappa$ is not coupled to the memory kernel; the term multiplied by $\kappa_{\rm r}$ represents the competition of the current density associated with a collision-induced reaction and the diffusive flux during momentum relaxation. 
We showed that the current density is reduced by the positive correlation between the velocity vector and the inward normal component of the velocity vector at the contact distance.  
Without $\kappa_{\rm r}$, the reaction rate coefficient is overestimated. 
Equation (\ref{eq:formalsolP9}) can be rewritten in the form of a generalized reaction--telegraph equation [Eq. (\ref{eq:2_28})], where a reaction--telegraph equation is generalized to include the reduction effect of the current density by the positive correlation of the current density at the contact distance. 
We also derived the same equation as Eq. (\ref{eq:formalsolP9}) by introducing decoupling between the configurational distribution and the momentum distribution, where an equilibrium Maxwell (Gaussian) distribution is assumed for the momentum. 

The long-time rate coefficient turned out to be influenced by $\kappa_{\rm r}$; the intrinsic reaction rate constant in the expression of the long-time rate coefficient reduces to $2/3$ of the original value for a collision-induced reaction without reflection at the contact distance in the lowest order of the perturbation expansion for the reaction sink term and the streaming term.
The persistent inertial effects that reduce the reaction rate coefficient are consistent with the results of Langevin dynamic simulations. \cite{Yang_01} Moreover, we examined the long-time rate coefficient by changing the momentum relaxation time denoted by $\tau_{\rm D}$. 
The diffusion constant is proportional to $\tau_{\rm D}$, whereas the reaction rate constants such as $\kappa$ and $\kappa_{\rm r}$ are independent of $\tau_{\rm D}$. 
When $\tau_{\rm D}$ is small, the long-time rate coefficient can be approximated by the diffusion controlled rate, which is given by $4 \pi \sigma D$ and is proportional to the contact distance. 
When $\tau_{\rm D}$ is increased,  the long-time rate coefficient is influenced by the rate constant given by 
$\kappa=2\sigma^2\sqrt{2\pi k_{\rm B} T/\mu}$ [Eq. (\ref{eq:28})] and 
is proportional to the square of the contact distance; in the reaction-controlled limit, reactions proceed by ballistic collisions. 

For electron transfer and energy transfer, the intrinsic reaction rate constant in the reaction sink term of the Fokker--Planck--Kramers equation can be independent of the reactant velocity. 
When the intrinsic reaction rate constant is independent of the reactant velocity and is localized, we have $\kappa_{\rm r}=0$; the aforementioned reduction of the long-time rate coefficient is absent in this case.  

For brevity, we ignored the liquid structure factor and the hydrodynamic effect, which might influence reaction kinetics. \cite{Kapral_78,Shin_78,Zhou_91,Lee_04,Northrup_79}
The hydrodynamic effect can be taken into account by the relative distance dependence in the diffusion constant. 
The structure factor can be taken into account by introducing the potential of mean force. 
Although we considered a simplified model, we found that the inertial effect reduces the long-time reaction rate coefficient for a collision-induced reaction.

\begin{acknowledgments}
% Put your acknowledgments here.
SL notes that this work was supported by the National Research Foundation of Korea (NRF) grant funded by the Korea government (MSIT) (No. 2020R1F1A1071933).
S.D.T. notes that this work was performed in accordance with
STATE TASK 45.12 (Grant No. 0082-2019-0009), registration number
in GZ 122040500058-1. Scientific basis for the designing of new
materials with desired properties and functions, including highpurity
and nanomaterials. Topic 1.1 “Physics and chemistry of new
nanostructured systems and composite materials with prescribed
properties.”
\end{acknowledgments}
%\newpage
\renewcommand{\theequation}{A.\arabic{equation}}  
\setcounter{equation}{0}  % reset counter     
\section*{Appendix A. Derivation of Eq. (\ref{eq:2})}
%\section*{Appendix A. }
\label{sec:AppendixA}

The Fokker--Planck--Kramers equation without the reaction sink term must be conservative, and we infer 
\begin{align}
	\int_{r\geq \sigma -\epsilon} d \bm{r} \int d \bm{v}\, \left[
	-\frac{\partial}{\partial \bm{r}} \cdot \bm{v}+\frac{\partial}{\partial \bm{v}} \frac{1}{\mu} \cdot \frac{\partial U}{\partial \bm{r}}+
	\frac{\partial}{\partial \bm{v}} \frac{1}{\tau_{\rm D}} \cdot \left(\bm{v} +\frac{k_{\rm B} T}{\mu} \frac{\partial}{\partial \bm{v}}
	\right)
	\right] f(\bm{r}, \bm{v},t) =0, 
	\label{eq:cons}
\end{align}
where a reflecting boundary condition is set at $r=\sigma -\epsilon$ by assuming isotropy.
%\sout{If} 
Provided function $f(\bm{r}, \bm{v},t)$ vanishes exponentially as $|\bm{v}| \rightarrow \infty$, we arrive at 
\begin{align}
	\int d \bm{v}\, 
	\frac{\partial}{\partial \bm{v}} \frac{1}{\tau_{\rm D}} \cdot \left(\bm{v} +\frac{k_{\rm B} T}{\mu} \frac{\partial}{\partial \bm{v}}
	\right) f(\bm{r}, \bm{v},t) =0,  
	\label{eq:cons1}\\
	\int d \bm{v}\, \frac{\partial}{\partial \bm{v}} \frac{1}{\mu} \cdot \frac{\partial U}{\partial \bm{r}} f(\bm{r}, \bm{v},t) =0 .
\end{align}
We then must have 
\begin{align}
	\int_{r\geq \sigma -\epsilon}  d \bm{r} \int d \bm{v}\, 
	\frac{\partial}{\partial \bm{r}} \cdot \bm{v} f(\bm{r}, \bm{v},t) =0 . 
	\label{eq:cons2}
\end{align}
We assume the initial equilibrium reactant distribution. 
Using the divergence 
%\sout{Gauss's} 
theorem and assuming that the influence of a reaction on the current density at the infinity distance is negligible, 
\begin{align}
	-\lim_{r\rightarrow \infty} \int d \bm{v}\, 
	\bm{n}\cdot \bm{v}f(\bm{r}, \bm{v},t) =0 ,  
	\label{eq:cons3}
\end{align}
we obtain
\begin{align}
	\left. \int d \bm{v}\, 
	\bm{n}\cdot \bm{v}f(\bm{r}, \bm{v},t) \right|_{r=\sigma-\epsilon}=0 ,
	\label{eq:cons3_1}
\end{align}
for the isotropic distribution of reactants. 

%%%%%%%%%%%%%%%%%%%%%%%%%%%%%%%%%%%%%%%%%%%%%%%%%%%%%%%%%%%%%%%%%%%%%%%%%%%%%
\section*{Appendix B. Derivation of the reaction rate coefficient [Eq. (\ref{eq:ratec1})]}
%\section*{Appendix A. }
\label{sec:AppendixB}
We consider bimolecular reaction between the reactants denoted by A and the reactants denoted by B, 
where the volume number density of B denoted by $c_{\rm B}$ is larger than the volume number density of A.
The survival probability of A is denoted by $S(t)$, which is the probability of finding A survived from the bimolecular reaction with B at time $t$. 
We also introduce the number of B reactants in the spherical volume ($V$) by $N_{\rm B} =c_{\rm B} V$, where  
the volume is expressed as $V=4 \pi [r_{\rm o}^3-(\sigma-\epsilon]^3)/3$ using $r_{\rm o}$ representing the distance to an outer-spherical boundary from the center of A. 
Note that the pair correlation function approaches $1$ as $r \rightarrow \infty$;  
$r_{\rm o}$ should be large enough to ensure $ \int_{\sigma-\epsilon}^{r_{\rm o}} 4\pi r^2 dr\rho(r,0) \approx V$. 
When A is surrounded by B reactants, the survival probability of A decays by reaction to one of $N_{\rm B}$ B-reactants,  
which can be expressed as 
\begin{align}
S(t)=\left(\int_{\sigma-\epsilon}^{r_{\rm o}} 4\pi r^2 dr\rho(r,t)/ \int_{\sigma-\epsilon}^{r_{\rm o}} 4\pi r^2 dr\rho(r,0)
\right)^{N_{\rm B}} , 
\label{eq:suvA}
\end{align} 
where the pair correlation function associated with each B reactant is assumed to be independent and identical, 
therefore is given by using the same expression, $\rho(r,t)$. 
From the definition, $S(0)=1$ is satisfied.   
By taking the thermodynamic limit, 
$S(t)$ is related to $\rho(r,t)$ as \cite{TACHIYA_83}
\begin{align}
S(t) &=  \lim_{r_{\rm o} \rightarrow \infty}  
\left(1-\frac{1}{V} \int_{\sigma-\epsilon}^{r_{\rm o}} 4\pi r^2 dr \left[\rho(r,0)-\rho(r,t)\right] 
\right)^{c_{\rm B}V}
\label{eq:Sv0}
\\
&= \exp\left[-c_{\rm B}  \int_{\sigma-\epsilon}^{\infty} 4\pi r^2 dr \left[\rho(r,0)-\rho(r,t)\right]
\right] , 
\label{eq:Sv}
\end{align}  
where $S(0)=1$ can be confirmed. 
Note that the original derivation is formulated in terms of the survival probability of B; \cite{TACHIYA_83} obviously,  the derivation can be directly applied to the pair correlation function. \cite{Rosspeintner_07}
The first-order reaction rate coefficient of A can be given by, 
\begin{align}
k_{\rm s}(t) = -\left[\frac{d}{dt} S(t) \right]/S(t)=-c_{\rm B} \int_{\sigma}^{\infty} 4\pi r^2 dr \frac{d}{dt} \rho(r,t)=c_{\rm B}\frac{d}{dt} p(t). 
\label{eq:ks}
\end{align}  
Here, we consider the second-order reaction rate coefficient given by $k(t)=k_{\rm s}(t)/c_{\rm B}$.


%%%%%%%%%%%%%%%%%%%%%%%%%%%%%%%%%%%%%%%%%%%%%%%%%%%%%%%%%%%%%%%%%%%%%%%%%%%%%%%%%%%%%%

\end{document}